\documentclass[11pt]{article}
\usepackage[totalwidth=385pt,totalheight=599pt]{geometry}
\usepackage{amsmath,amssymb,epsf,graphics,graphicx}
\usepackage{bbm}
\newcommand\R[2]{R_{#1}^{#2}}
\newcommand{\One}{\mathbbm{1}}

\newcommand{\rl}{\rule[-0.5cm]{0cm}{1.cm}}
\newcommand{\rll}{\rule[-0.5cm]{0cm}{1.2cm}}

\newcommand{\usa}[1]{ \left \{ #1 \right \} }
\newcommand{\ubl}[1]{\{#1\}}
\newcommand{\bl}[1]{\left(#1\right)}
\newcommand{\blk}[1]{\langle #1\rangle}
\newcommand{\tbl}[1]{\langle\widetilde{#1}\rangle}
\newcommand\blank[1]{}

\newcommand{\fract}[2]{{\textstyle\frac{#1}{#2}}}

\newcommand{\ket}[1]{\ensuremath{\mbox{\normalsize $| #1\rangle$}}}
\newcommand{\sket}[1]{| #1\rangle}

\newcommand{\sbk}{\langle \alpha | \alpha\rangle}
\newcommand{\sbkp}{\langle + | +\rangle}
\newcommand{\sbAB}[2]{\langle #1 | #2 \rangle}
\newcommand{\bra}[1]{\ensuremath{\mbox{\normalsize $\langle #1|$}}}
\newcommand{\vev}[1]{\langle\,#1\,\rangle}
\renewcommand{\hat}{\widehat}
\newcommand\eq{\begin{equation}}
\newcommand\en{\end{equation}}
\newcommand\bea{\begin{eqnarray}}
\newcommand\eea{\end{eqnarray}}
\newcommand\nn{\nonumber}
\newcommand\ba{\(\begin{array}}
\newcommand\ea{\end{array}\)}
\newcommand{\resection}[1]{\setcounter{equation}{0}\section{#1}}

\newcommand{\CG}{{\cal G}}
\newcommand{\Z}{{\mathbb Z}}
\newcommand{\NN}{{\mathbb N}}

\newcommand\te{\theta}
\newcommand\bzero{\boldsymbol{0}}
\newcommand\bmu{\boldsymbol{\mu}}
\newcommand\bnu{\boldsymbol{\nu}}
\newcommand\brho{\boldsymbol{\rho}}
\newcommand\bLambda{\boldsymbol{\Lambda}}
\newcommand\blambda{\boldsymbol{\lambda}}
\newcommand\rg{r_{\mathfrak{g}}}
\newcommand\Rth{{\mathbb R}}
\newcommand\ep{\varepsilon}
\newcommand{\ivec}[1]{|\,#1\,\rangle\!\rangle}
\begin{document}
\begin{titlepage}
\vskip 0.5cm
\begin{flushright}
DCPT-05/63 \\
{\tt hep-th/0512337}
\end{flushright}
\vskip .7cm
\begin{center}
{\Large{\bf Reflection factors and exact $g$-functions\\[3pt]
for purely elastic scattering theories}}
\end{center}
\vskip 0.8cm
\centerline{Patrick Dorey$^1$, Anna Lishman$^1$,
Chaiho Rim$^{2}$ and Roberto Tateo$^{3}$}
\vskip 0.9cm
\centerline{${}^1$\sl\small Dept.\ of Mathematical Sciences,
University of Durham,}
\centerline{\sl\small  Durham DH1 3LE, United Kingdom\,}
\vskip 0.3cm
\centerline{${}^{2}$\sl\small Dept.\ of Physics and Research Institute
of Physics and Chemistry,}
\centerline{\sl\small
Chonbuk National University,
Chonju 561-756, Korea}
\vskip 0.2cm
\centerline{${}^{3}$\sl\small Dip.\ di Fisica Teorica and INFN,
Universit\`a di Torino,}
\centerline{\sl\small Via P.\ Giuria 1, 10125 Torino, Italy}
\vskip 0.2cm
\centerline{E-mails:}
\centerline{p.e.dorey@durham.ac.uk, a.r.lishman@durham.ac.uk,}
\centerline{rim@chonbuk.ac.kr,  tateo@to.infn.it}

\vskip 1.25cm
\begin{abstract}
\noindent
We discuss reflection factors for purely elastic scattering theories
and relate them to perturbations of specific conformal boundary
conditions, using recent results on exact off-critical $g$-functions.
For the non-unitary cases, we support our conjectures using a
relationship with quantum group reductions of the sine-Gordon model.
Our results imply the existence of a variety of new flows
between conformal boundary conditions, some of them driven by
boundary-changing operators.
\end{abstract}
\end{titlepage}
\setcounter{footnote}{0}
\def\thefootnote{\fnsymbol{footnote}}
%
\resection{Introduction}
There are two general approaches to the study of integrable
boundary quantum field theories: infrared and ultraviolet. From the
infrared point of view, solving the  boundary Yang-Baxter and
bootstrap constraints allows sets of reflection factors to be
associated with various bulk scattering theories.  However, there
are many ambiguities left by these constraints, which cannot be
lifted without the introduction of more selective criteria. This is
to be expected, since a given bulk theory may admit many different
integrable boundary conditions.

Alternatively, a theory
might be defined via a classical Lagrangian,
or as a perturbation of a boundary conformal field theory. In
either case, it is natural to ask for a `UV/IR dictionary', giving
the relationship between the infrared and ultraviolet specifications
of the model. In particular, this would reveal
which among the infinite families of solutions to the boundary
Yang-Baxter and bootstrap equations are physically realised, and
to which boundary conditions they correspond.

Progress on this issue includes the relationship between the
parameters  in the boundary reflection factors  and the couplings in
the boundary sine-Gordon Lagrangian  derived by Aliosha
Zamolodchikov~\cite{bologna} (checked using the boundary TCSA
\cite{Dorey:1997yg} in \cite{Bajnok:2001ug}),
and various perturbative results for
cases where the theory admits a Lagrangian definition (see for
example \cite{Corrigan:2000fm,Kormos:2002ya}). However, these
methods cannot be readily adapted to the study of general perturbed
boundary conformal field theories. In particular, a boundary analogue
of the thermodynamic Bethe ansatz (TBA) $c$-function, which allows
bulk S-matrices to be identified with specific perturbed
conformal field theories
\cite{Zamolodchikov:1989cf,Klassen:1989ui},
has been lacking.
An obvious candidate is the $g$-function defined by  Affleck and
Ludwig in \cite{Affleck:1991tk}, but the initial proposal for a
TBA-like equation for a fully off-critical version of $g$, taking as input
just the bulk S-matrix and the boundary reflection factors
\cite{LeClair:1995uf}, turned out to be incorrect
\cite{Dorey:1999cj}.

Recently it has been shown that the situation can be
remedied\,\cite{Dorey:2004xk}. The detailed work in
\cite{Dorey:2004xk} concentrated on the boundary
scaling Lee-Yang model, for which the relationship between
microscopic boundary conditions and reflection factors had already
been established\,\cite{Dorey:1997yg}.
In this paper we extend the investigations of
\cite{Dorey:2004xk} to a collection of theories for which boundary
UV/IR relations have yet to be found, namely
the minimal purely elastic scattering theories associated with the
ADET series of diagrams
\cite{Braden:1989bu,Klassen:1989ui,Fateev:1990hy,Dorey:1990xa,%
Dorey:1991zp,Fring:1991gh,Ravanini:1992fi}.
The bulk S-matrices of these models have long been known, but less
progress has been made in associating solutions of the boundary
bootstrap equations with specific perturbed boundary conditions.  We
present a collection of minimal reflection factors for the ADET
theories, and test them by checking the $g$-function flows that they
imply. We also show how these reflection factors can be modified to
incorporate a free parameter, generalising a structure previously
observed in the Lee-Yang model
\cite{Dorey:1997yg}. This enables us to predict a number of new
flows between conformal boundary conditions.

In more detail, the rest of the paper
is organised as follows. Bulk and boundary
ADET scattering theories are discussed in sections 2 and 3, with
section 3 containing a full set of reflection
factors fulfilling certain minimality criteria. These
solutions have no free parameters, and in section 4 the minimality
requirement is relaxed and non-minimal one-parameter families of
solutions are proposed to describe simultaneous bulk and boundary
perturbations of boundary
conformal field theories by relevant operators.
In section 5 the one-parameter families of reflection amplitudes
for the $T_r$ models are alternatively deduced
by considering special reductions of the
boundary sine-Gordon model.

After a discussion of the exact off-critical $g$-function in section
6, the physical consistency of the  minimal  solutions is checked
in section~7, by comparing the predictions for the
ultraviolet values of the $g$-function obtained from the reflection
factors with conformal field theory results.
An additional check in first order perturbation theory is
performed for the 3-state Potts model in section~8.  Section~9
summarises the $g$-function predictions for the
one-parameter families of reflection factors and discusses their UV/IR
relations. Our
conclusions are given in section~10, and an appendix gathers together
various results concerning the fields and $g$-function values in the
diagonal
$\hat{\mathfrak{g}}_1\times\hat{\mathfrak{g}}_1/\hat{\mathfrak{g}}_2$
coset models
which are
needed in the main text.

\resection{The ADET family of purely elastic scattering theories}
\label{ADETfS}

Purely elastic scattering theories are characterised by the property
that their S-matrices are diagonal. The scattering of particles $a$
and $b$ with relative rapidity $\theta$ is then described by a
single function $S_{ab}(\theta)$, which is a pure phase for physical
rapidities. The unitarity and crossing symmetry
conditions simplify to
\bea
S_{ab}(\theta)S_{ab}(-\theta)&=&1~,
\label{sunitarity}\\[3pt]
S_{ab}(\theta)&=&S_{a\bar b}(i\pi-\theta)
\label{scross}
\eea
respectively, where $\bar b$ is the antiparticle of $b$. The
Yang-Baxter equation is trivially satisfied, but, as stressed
by Zamolodchikov~\cite{Zamolodchikov:1989fp,Zamolodchikov:1990jh},
the bootstrap still provides a useful constraint. This states that
whenever an on-shell three-point coupling
$C^{abc}$ is nonzero, the S-matrix elements of particles $a$, $b$ and
$\bar c$ with any other particle $d$ satisfy
\eq
S_{d\bar c}(\theta)=S_{da}(\theta-i\bar U_{ac}^b)
S_{db}(\theta+i\bar U_{bc}^a)
\label{sboot}
\en
where the `fusing angles' $U_{ab}^c$ are related to the masses of
particles $a$, $b$ and $c$ by
\eq
M_c^{2}=M_a^{2}+M_b^{2}+2M_a M_b\cos U^c_{ab}
\label{fusdef}
\en
and $\bar U=\pi-U$. These angles also appear as the locations of
certain odd-order poles in the S-matrix elements. More details of
the workings of the bootstrap can be found
in~\cite{Zamolodchikov:1989fp,Zamolodchikov:1990jh} and, for
example, \cite{Dorey:1996gd}.

The purely elastic scattering theories that we shall treat in this
paper fall into two classes. The first class associates an
S-matrix to each simply-laced Lie algebra $\mathfrak{g}$, of type A, D or E
\cite{Braden:1989bu,Klassen:1989ui,Fateev:1990hy,Dorey:1990xa}.
These S-matrices are {\em minimal}, in that they have no zeros on
the physical strip $0\le \Im m\,\theta\le \pi$, and {\em
one-particle unitary}, in that all on-shell three-point couplings,
as inferred from the  residues of forward-channel S-matrix poles,
are real. They describe particle scattering in the
perturbations of the coset conformal field theories
$\hat{\mathfrak{g}}_1\times \hat{\mathfrak{g}}_1 /\hat{\mathfrak{g}}_2$
by their $(1,1,\mbox{ad})$ operators, where $\hat{\mathfrak{g}}$ is the affine
algebra associated with $\mathfrak{g}$\,. The unperturbed theories have central
charge
\eq
c= \frac{2 r_{\mathfrak{g}}}{(h+2)}~,
\en
where $\rg$ is the rank of $\mathfrak{g}$, and
$h$ is its Coxeter number. These UV central charges can be recovered
directly from the S-matrices, using the thermodynamic Bethe
ansatz~\cite{Zamolodchikov:1989cf,Klassen:1989ui}.

The ADE S-matrices describe the diagonal scattering of $\rg$
particle types, whose masses together form
the Perron-Frobenius eigenvector of the Cartan matrix of
$\mathfrak{g}$. This allows the particles to be attached to the nodes
of the Dynkin diagram of $\mathfrak{g}$.
Each S-matrix element can be conveniently written as a
product of elementary blocks\,\cite{Braden:1989bu}
\eq
\usa{x} = \bl{x-1}\bl{x+1}~,\quad
\bl{x}(\theta) =
 \frac{\sinh\left(\frac{\theta}{2}+\frac{i\pi x}{2h}\right)}
   {\sinh\left(\frac{\theta}{2}-\frac{i\pi x}{2h}\right)}
\label{bblock}
\en
as
\eq
S_{ab}=\prod_{x\in A_{ab}}\usa{x},
\label{smat}
\en
for some index set $A_{ab}$.
Note that
\eq
\bl{0}=1\,,\quad
\bl{h}=-1\,,\quad
\bl{-x}=\bl{x}^{-1},\quad
\bl{x\pm 2h}=\bl{x}.
\label{blprops}
\en
The notation (\ref{bblock}) has been
so arranged that the numbers $x$ are all integers. The sets
$A_{ab}$ are tabulated in \cite{Braden:1989bu}; a universal
formula expressing them in geometrical terms was found
in~\cite{Dorey:1990xa}, and is further discussed
in \cite{Dorey:1991zp,Fring:1991gh}.

The S-matrices of the second class
\cite{Smirnov:1989hh,Cardy:1989fw,Freund:1989jq} are
labelled by extending the set of ADE Dynkin diagrams to include the
`tadpole' $T_r$. They encode the diagonal scattering of $r$
particle types, and can be written in terms of the blocks (\ref{bblock})
with $h=2r{+}1$ \cite{Dorey:1997rb}\,:
\eq
S_{ab}=
\prod_{\substack{l=|a-b|+1 \\ \textrm{step 2}}}^{a+b-1}\{l\}\{h-l\}\,.
\label{trS}
\en
These S-matrix elements are again minimal, but they are not
one-particle unitary, reflecting the fact that they
describe perturbations of the non-unitary minimal models ${\cal
M}_{2,2r{+}3}$, with central charge
$c=-2r(6r+5)/(2r+3)$. The perturbing operator this time is
$\phi_{13}$\,. The $T_r$ S-matrices are quantum group
reductions of the sine-Gordon model at coupling
$\beta^{2}=16\pi/(2r{+}3)$ \cite{Smirnov:1989hh}, a fact
that will be relevant later.

A self-contained classification of minimal purely elastic S-matrices
is still lacking, but the results of~\cite{Ravanini:1992fi} single
out the ADET theories as the only examples having TBA systems for which
all pseudoenergies remain finite in the ultraviolet limit.

The  ADET diagrams are shown in figure~\ref{fig1}, with nodes
giving our conventions for labelling the particles in each theory.
For the $D_r$ theories, particles $r{-}1$ and $r$ are sometimes
labelled $s$ and $s'$, or $s$ and $\bar s$, for $r$ even or odd
respectively.
\[\begin{array}{ccc}
\hskip-15pt
\epsfxsize=.27\linewidth\epsfbox{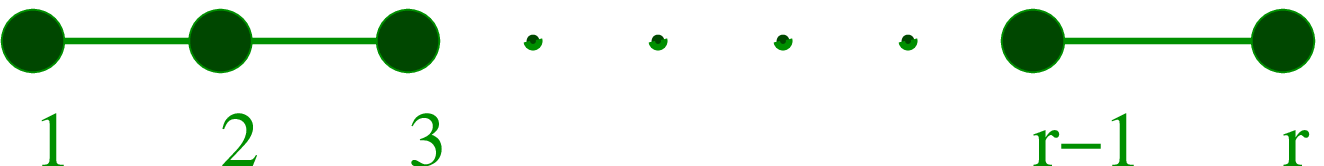} {}~~~&
\epsfxsize=.27\linewidth\epsfbox{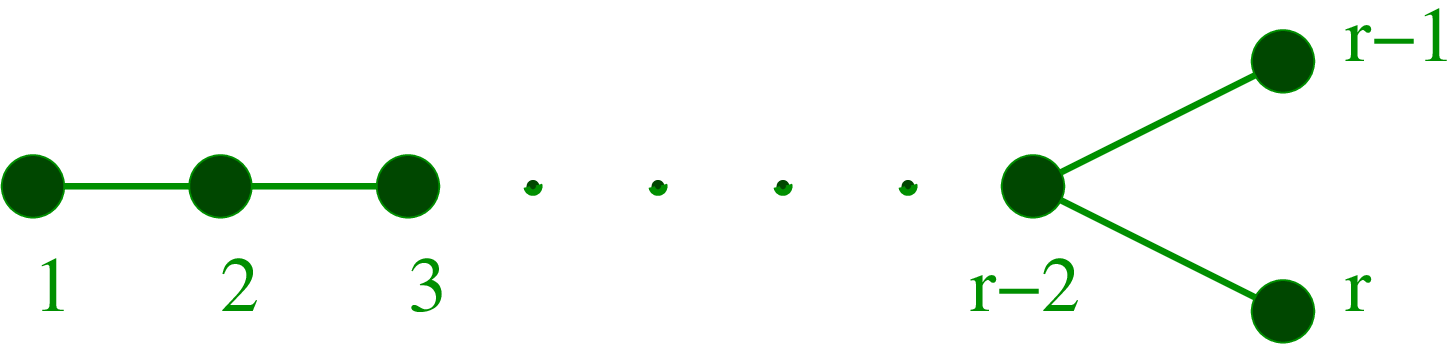} & {}~~~~
\epsfxsize=.27\linewidth\epsfbox{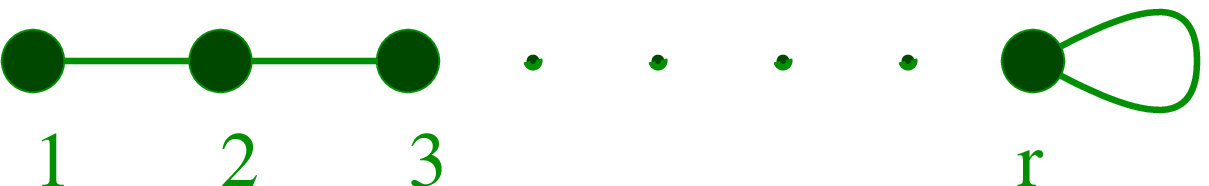}
\\[5pt]
\parbox[t]{.2\linewidth}{\quad~~~\small Fig.~\protect\ref{fig1}a: $A_r$}
{}~~&~
\parbox[t]{.2\linewidth}{\quad~~\small  Fig.~\protect\ref{fig1}b: $D_r$}
{}~~&~
\parbox[t]{.2\linewidth}{\quad~~\small  Fig.~\protect\ref{fig1}c: $T_r$}
\\[20pt]
\hskip-30pt
\epsfxsize=.18\linewidth\epsfbox{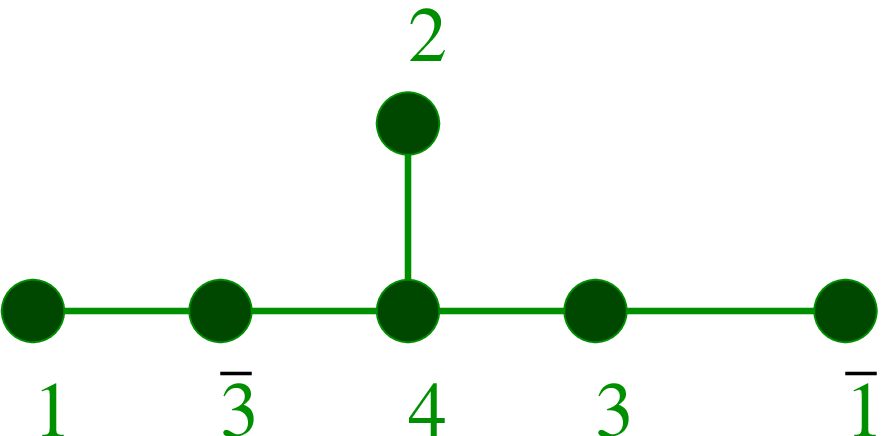} {}~&
\epsfxsize=.21\linewidth\epsfbox{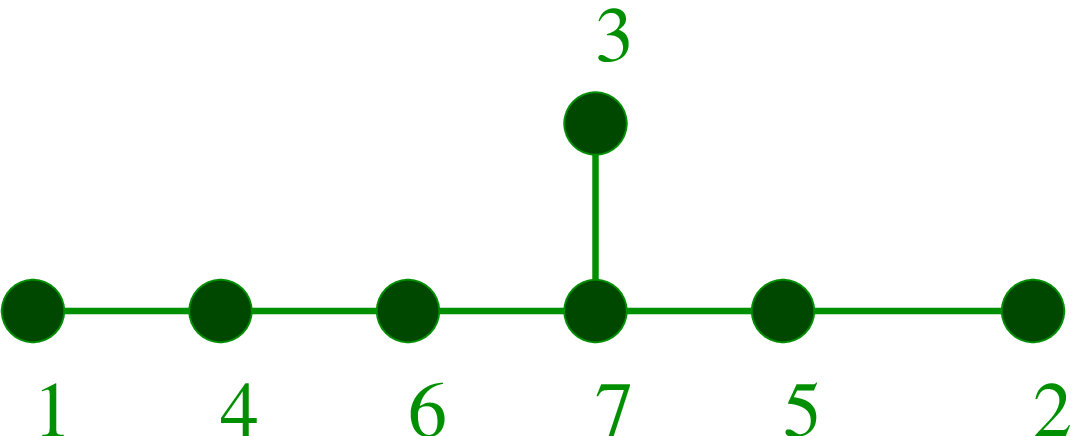} ~&
\epsfxsize=.24\linewidth\epsfbox{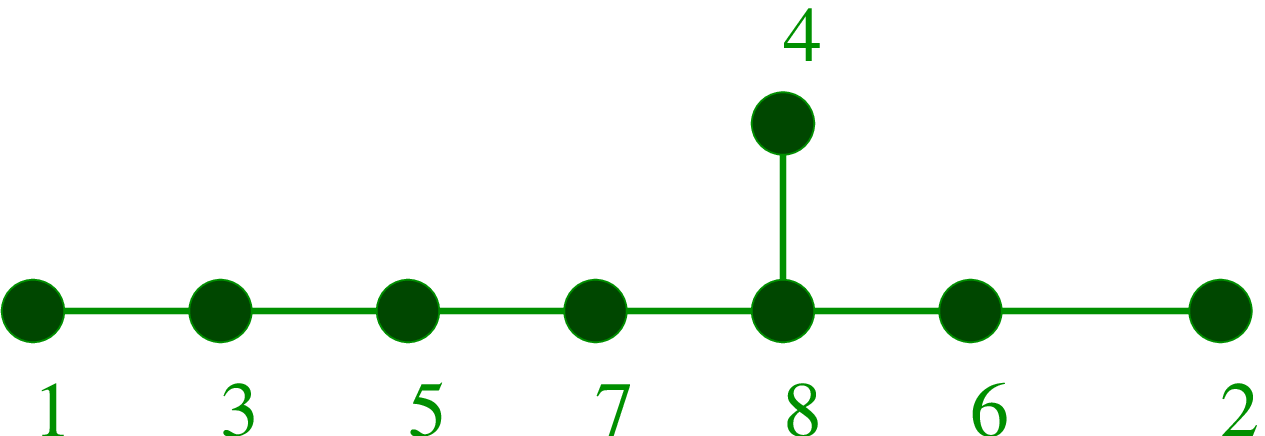}
\\
\parbox[t]{.2\linewidth}{\quad~~~\small Fig.~\protect\ref{fig1}d:
$E_6$}
 {}~~&~
\parbox[t]{.2\linewidth}{\quad~~\small  Fig.~\protect\ref{fig1}e: $E_7$}
 {}~~&~
\parbox[t]{.2\linewidth}{\quad~~\small  Fig.~\protect\ref{fig1}f: $E_8$}
\\[5pt]
\end{array}
\refstepcounter{figure}
\protect\label{fig1}
\]

\resection{Minimal reflection factors for purely elastic scattering
theories}
\label{sec.R}
The scattering of particles by a boundary in an integrable quantum
field theory is described by a set of reflection factors. In this paper
we shall restrict ourselves to situations where not only the bulk, but
also the boundary scattering amplitudes are purely elastic.
The reflection factors are then rapidity-dependent functions
$\R{a}{}(\theta)$, one for each particle type $a$ in the theory.
Unitarity \cite{Fring:1993mp,Ghoshal:1993tm}
and crossing-unitarity \cite{Ghoshal:1993tm}
constrain these functions as
\bea
\R{a}{}(\theta)\R{a}{}(-\theta)&=&1\,,
\label{unitarity}\\[3pt]
\R{a}{}(\theta)\R{\bar a}{}(\theta-i\pi)&=&S_{aa}(2\theta)\,.
\label{crossunitarity}
\eea
In addition, whenever a bulk three-point coupling $C^{abc}$ is
nonzero, a boundary bootstrap equation holds
\cite{Fring:1993mp,Ghoshal:1993tm}:
\eq
\R{\bar c}{}(\theta)= \R{a}{}(\theta+i U^b_{ac})
\R{b}{}(\theta-i\bar U^a_{bc}) S_{ab}(2\theta+i\bar U^b_{ac}-i\bar
U^a_{bc})\,.
\label{bboot}
\en
For a given bulk S-matrix there are infinitely-many distinct sets of
reflection factors consistent with these constraints, because
any solution can be multiplied by a solution of the bulk bootstrap,
unitarity and crossing equations to yield another solution
\cite{Sasaki:1993xr}.
To identify a set of reflection factors as physically relevant some
more information is needed. A common basis for the conjecturing of
bulk scattering amplitudes is a `minimality hypothesis', that in the
absence of other requirements one should look for solutions of the
constraints with the smallest possible number of poles and zeros. In
this section we show how the same principle can be used to find sets
of boundary amplitudes, one for each purely elastic S-matrix of type
A, D or E. The amplitudes given below were at first
conjectured~\cite{doreyup} as a natural generalisation of the
minimal versions of the
$A_r$ affine Toda field theory amplitudes found in
\cite{Corrigan:1994ft,Corrigan:1994np}. The reasoning behind these
conjectures will be explained shortly. More recently, Fateev
\cite{Fateev:2001mj} proposed
a set of reflection factors for the affine Toda
field theories. These were in integral form, and not all
matrix elements were given. However, modulo some overall signs and
small typos, the coupling-independent parts of
Fateev's conjectures match ours. (This has also been found by Zambon
\cite{cz}, who obtained equivalent formulae to those recorded below
taking Fateev's integral formulae as a starting point.)

Unitarity and crossing-unitarity together imply that the
reflection factors must be $2\pi i$ periodic; unitarity then requires
that they be products of the blocks $(x)$ introduced in the last
section. The crossing-unitarity constraint is then key for the
analysis of minimality.  Each pole or zero of $S_{aa}(2\theta)$ on
the right hand side of (\ref{crossunitarity}) must be present in one
or other of the factors on the left hand side of that equation. This
sets a lower bound on the
number of poles and zeros for the reflection factor
$\R{a}{}(\theta)$.

To exploit this observation, it is convenient to work at the
level of the larger
blocks $\{x\}$, the basic
units for the bulk bootstrap equations
\cite{Braden:1989bu,Dorey:1990xa}. We define
two complementary `square roots' of these blocks as
\eq
\blk{x} =
\left(\fract{x-1}{2}\right)\left(\fract{2h-x-1}{2}\right)^{-1}~~,~~~
\tbl{x} = \left(\fract{x+1}{2}\right)\left(\fract{2h-x+1}{2}\right)^{-1}
\en
and record their basic properties
\eq
\blk{x}(\theta) = \tbl{x}(\theta+i\pi)
\en
and
\eq
\blk{x}(\theta)\,\tbl{x}(\theta)=\usa{x}(2\theta)\,.
\en
The crossing-unitarity equation can then be solved, in
a minimal fashion, by any product
\eq
\R{a}{}=\prod_{x\in A_{aa}}f_x
\label{Rform}
\en
where each factor $f_x$ can be freely chosen to be $\blk{x}$
or $\tbl{x}$, modulo one subtlety:
since $(0)=1$, minimality requires that
any factor $f_1$ be taken to be $\blk{1}$ rather
than $\tbl{1}$. In fact there is always exactly one such factor for
the diagonal S-matrix elements relevant here
\cite{Dorey:1991zp}.\footnote{We could also set $\blk{x} =
\left(\fract{x-1}{2}\right)\left(\fract{x+1}{2}\right)$,
$\tbl{x} =
\left(\fract{2h-x-1}{2}\right)^{-1}\left(\fract{2h-x+1}{2}\right)^{-1}$
throughout,
but this would break the pattern previously seen for the $A_r$ theories,
and turns out not to fit with the $g$-function calculations later.}

There remain the bootstrap equations (\ref{bboot}).
To treat these we need to know how the blocks
$\blk{x}(\theta)$ and $\tbl{x}(\theta)$ transform under general shifts
in $\theta$. As in
\cite{Dorey:1990xa,Dorey:1991zp}, these shifts are best discussed by
defining
\eq
\bl{x}_+(\theta) =
 \sinh\left(\frac{\theta}{2}+\frac{i\pi x}{2h}\right)
\en
and then
\eq
\blk{x}_+ =
\left(\fract{x-1}{2}\right)_+\left(\fract{2h-x-1}{2}\right)_+^{-1}~~,~~~
\tbl{x}_+ =
\left(\fract{x+1}{2}\right)_+\left(\fract{2h-x+1}{2}\right)_+^{-1}
\en
so that $(x)=(x)_+/(-x)_+$,
$\blk{x}=\blk{x}_+/\tbl{-x}_+$ and
$\tbl{x}=\tbl{x}_+/\blk{-x}_+$\,.
Once a putative set of reflection factors
(\ref{Rform}) has been decomposed into these blocks, it is
straightforward to implement the boundary bootstrap equations
(\ref{bboot}) using the properties
\eq
\blk{x}_+(\theta+i\pi y/h) =\blk{x{+}y}_+(\theta)~,~~
\tbl{x}_+(\theta+i\pi y/h) =\tbl{x{+}y}_+(\theta)\,.
\en

For the ADE theories
it turns out that the boundary bootstrap equations {\em can\/}
all be satisfied by minimal conjectures of the form (\ref{Rform}),
and that the choice of blocks is then fixed uniquely.
(From the point of view of the bootstrap equations alone, an overall swap
between $\blk{x}$ and $\tbl{x}$ is possible, but the requirement that
$f_1=\blk{1}$ fixes even this ambiguity.)
The final answers
are recorded below.
\subsection*{$A_r$}
The reflection factors for this case were given in
\cite{Corrigan:1994ft}. In the current notation they are
\eq
\R{a}{} = \prod_{\substack{x=1 \\ \textrm{$x$ odd}}}^{2a-1}
 \blk{x}
\en

\subsection*{$D_r$}
The reflection factors, $\R{a}{}$, for $a=1,\ldots,r-2$ are
\bea
\R{a}{} &=& \prod_{\substack{x=1 \\ \textrm{$x$ odd}}}^{2a-1}
\blk{x}
 \prod_{\substack{x=1 \\ \textrm{step 4}}}^{2a-1} \blk{h-x}
 \prod_{\substack{x=1 \\ \textrm{step 4}}}^{2a-5} \tbl{h-x-2}
 \, ,\,\textrm{for $a$ odd}  \\
\R{a}{} &=& \prod_{\substack{x=1 \\ \textrm{$x$ odd}}}^{2a-1}
\blk{x}
 \prod_{\substack{x=1 \\ \textrm{step 4}}}^{2a-3} \tbl{h-x}\blk{h-x-2}
 \, ,\,\textrm{for $a$ even} \nn
\eea
while for $a=r-1$ and $r$ we have
\bea
\R{r-1}{} &=& \R{r}{} = \prod_{\substack{x=1 \\ \textrm{step
4}}}^{2r-5}
 \blk{x} \, ,\,\textrm{for $r$ odd}  \\
\R{r-1}{} &=& \R{r}{} = \prod_{\substack{x=1 \\ \textrm{step
4}}}^{2r-3}
 \blk{x} \, ,\,\textrm{for $r$ even.} \nn
\eea

\subsection*{$E_6$}
\bea
\R{1}{} &=& \R{\bar 1}{} \,=\, \blk{1}\blk{7} \nn \\
\R{2}{} &=& \blk{1}\blk{5}\blk{7}\tbl{11} \\
\R{3}{} &=& \R{\bar{3}}{} \,=\,
\blk{1}\blk{3}\blk{5}\blk{7}\tbl{7}\blk{9} \nn \\
\R{4}{} &=&
\blk{1}\blk{3}^{2}\blk{5}^{2}\tbl{5}\blk{7}^{2}\tbl{7}\blk{9}\tbl{9}
 \blk{11}\nn
\eea
\subsection*{$E_7$}
\bea
\R{1}{} &=& \blk{1}\blk{9}\blk{17} \nn \\
\R{2}{} &=& \blk{1}\blk{7}\blk{11}\tbl{17} \nn \\
\R{3}{} &=& \blk{1}\blk{5}\blk{7}\blk{9}\tbl{11}\blk{13}\blk{17} \nn \\
\R{4}{} &=& \blk{1}\blk{3}\blk{7}\blk{9}\tbl{9}\blk{11}\blk{15}\tbl{17}  \\
\R{5}{} &=&
\blk{1}\blk{3}\blk{5}\blk{7}\tbl{7}\blk{9}^{2}\blk{11}\tbl{11}\blk{13}
 \tbl{15}\blk{17} \nn \\
\R{6}{} &=&
\blk{1}\blk{3}\blk{5}^{2}\blk{7}\tbl{7}\blk{9}^{2}\tbl{9}\blk{11}\tbl{11}
 \blk{13}^{2}\tbl{15}\blk{17} \nn \\
\R{7}{} &=&
\blk{1}\blk{3}^{2}\blk{5}^{2}\tbl{5}\blk{7}^3\tbl{7}\blk{9}^{2}\tbl{9}^{2}
 \blk{11}^3\tbl{11}\blk{13}\tbl{13}^{2}\blk{15}^{2}\tbl{17} \nn
\eea
\subsection*{$E_8$}
\bea
\R{1}{} \!&=&\! \blk{1}\blk{11}\blk{19}\tbl{29} \nn \\
\R{2}{} \!&=&\!
\blk{1}\blk{7}\blk{11}\blk{13}\tbl{17}\blk{19}\blk{23}\tbl{29} \nn \\
\R{3}{} \!&=&\!
\blk{1}\blk{3}\blk{9}\blk{11}\tbl{11}\blk{13}\blk{17}\blk{19}\tbl{19}
 \blk{21}\tbl{27}\blk{29} \nn \\
\R{4}{} \!&=&\!
\blk{1}\blk{5}\blk{7}\blk{9}\blk{11}\tbl{11}\blk{13}\blk{15}\tbl{15}
 \blk{17}\blk{19}\tbl{19}\blk{21}\tbl{23}\blk{25}\blk{29} \nn \\
\R{5}{} \!&=&\!
\blk{1}\blk{3}\blk{5}\blk{7}\blk{9}\tbl{9}\blk{11}^{2}\tbl{11}\blk{13}
 \tbl{13}\blk{15}^{2}\blk{17}\tbl{17}\blk{19}^{2}\tbl{19}\blk{21}
 \tbl{21}
 \nn \\
 & &\blk{23}\tbl{25}\blk{27}\tbl{29}  \\
\R{6}{} \!&=&\!
\blk{1}\blk{3}\blk{5}\blk{7}\tbl{7}\blk{9}^{2}\blk{11}^{2}\tbl{11}
 \blk{13}^{2}\tbl{13}\blk{15}\tbl{15}\blk{17}^{2}\tbl{17}\blk{19}
 \tbl{19}^{2}\nn \\
 & &\blk{21}^{2}\blk{23}\tbl{23}\blk{25}\tbl{27}\blk{29} \nn \\
\R{7}{} \!&=&\!
\blk{1}\blk{3}\blk{5}^{2}\blk{7}^{2}\tbl{7}\blk{9}^{2}\tbl{9}\blk{11}^{2}
 \tbl{11}^{2}\blk{13}^3\tbl{13}\blk{15}^{2}\tbl{15}^{2}\blk{17}^3\tbl{17}
 \blk{19}^{2} \nn \\
 & & \tbl{19}^{2}
\blk{21}^{2}\tbl{21}\blk{23}\tbl{23}^{2}\blk{25}^{2}\tbl{27}\blk{29} \nn \\
\R{8}{} \!&=&\!
\blk{1}\blk{3}^{2}\blk{5}^{2}\tbl{5}\blk{7}^3\tbl{7}\blk{9}^3\tbl{9}^{2}
 \blk{11}^4\tbl{11}^{2}\blk{13}^3\tbl{13}^3\blk{15}^4\tbl{15}^{2}\blk{17}^3
 \tbl{17}^3 \nn \\
 & &\blk{19}^4\tbl{19}^{2}\blk{21}^{2}\tbl{21}^3\blk{23}^3\tbl{23}\blk{25}
 \tbl{25}^{2}\blk{27}^{2}\tbl{29} \nn
\eea

For the T series the story is different: it is {\em not}\/ possible to
satisfy the boundary bootstrap equations with a conjecture of the form
(\ref{Rform}). This means that the minimal reflection factors for
these models are forced by the bootstrap
to have extra poles and zeros beyond those
required by crossing-unitarity alone. Our general proposal will be
given in eq.~(\ref{r0def}) below, but the situation can be understood
using the
boundary $T_1$, or Lee-Yang, model: the minimal reflection factor
found in \cite{Dorey:1997yg} for the single particle in this model
bouncing off the $\ket{\One}$ boundary is
\eq
R^{\sket{\One}}=(\fract{1}{2})(\fract{3}{2})(\fract{4}{2})^{-1}~.
\label{r1}
\en
The simpler function $(\fract{1}{2})(\fract{4}{2})^{-1}$ would have
been enough to satisfy crossing-unitarity, but then the boundary
bootstrap would not have held, and so (\ref{r1}) really is a minimal
solution. This observation fits nicely with the $g$-function
calculations to be reported later: had the minimal reflection factors for
the $T_r$ theories fallen into the pattern seen for other models,
there would have been a mismatch between the predicted UV values of
the $g$-functions and the known values from conformal field
theory.

\resection{One-parameter families of reflection factors}
\label{sec.Rb}
The minimal reflection factors introduced in the last section
have no free parameters. However, combined
perturbations of a boundary conformal field theory by relevant
bulk and boundary
operators involve
a dimensionless quantity --
the ratio of the induced bulk and boundary scales --
on which the reflection factors would
be expected to depend. To describe such situations, we must drop the
minimality hypothesis, and extend our
conjectures.

A first observation, rephrasing that of \cite{Sasaki:1993xr},
is that given {\em any}\/ two
solutions $R_a(\theta)$ and $R'_a(\theta)$ of the
reflection unitarity, crossing-unitarity and bootstrap relations
(\ref{unitarity}), (\ref{crossunitarity}) and (\ref{bboot}), their
ratios $Z_a(\theta)\equiv
R_a(\theta)/R'_a(\theta)$ automatically solve one-index versions
of the bulk unitarity, crossing and bootstrap equations
(\ref{sunitarity}), (\ref{scross}) and (\ref{sboot}):
\eq
Z_{a}(\theta)Z_{a}(-\theta)=1~,~~~
Z_{a}(\theta)=Z_{\bar a}(i\pi-\theta)~,
\label{zunicross}
\en
\eq
Z_{\bar c}(\theta)=Z_{a}(\theta-i\bar U_{ac}^b)
Z_{b}(\theta+i\bar U_{bc}^a)~.
\label{zboot}
\en
The minimal
reflection factors $R_a(\theta)$
can therefore be used as multiplicative
`seeds' for more
general conjectures $R'_a(\theta)=R_a(\theta)/Z_a(\theta)$,
with the parameter-dependent parts
$Z_a(\theta)^{-1}$ constrained via
(\ref{zunicross}) and (\ref{zboot}).
An immediate solution is
$Z^{\sket{d}}_a(\theta)=S_{da}(\theta)$
for any (fixed) particle type $d$ in the theory,
where we have used the ket symbol to indicate that the label
$\ket{d}$ might ultimately refer to one of the possible boundary
states of the theory. However, this does not yet introduce a parameter.
Noting that a symmetrical shift in $\theta$ preserves all the relevant
equations, one possibility is to take
\eq
Z^{\sket{d,C}}_a(\theta)=S_{da}(\theta+C)S_{da}(\theta-C)\,,
\label{zone}
\en
with $C$ at this stage arbitrary. This is indeed the solution
adopted by the boundary  scaling Lee-Yang example studied
in~\cite{Dorey:1997yg}. This model is the $r=1$ member of the $T_r$
series described earlier, and corresponds to the perturbation of the
non-unitary minimal model ${\cal M}_{25}$ by its only relevant bulk
operator, $\varphi$, of conformal dimensions
$\Delta_{\varphi}=\overline{\Delta}_{\varphi}=-\fract{1}{5}$. The
minimal model has two conformally-invariant boundary
conditions which were labelled
$\ket{\One}$ and $\ket{\Phi}$ in \cite{Dorey:1997yg}.
The
$\ket{\One}$ boundary has no relevant boundary fields,
and has a minimal reflection factor, given by (\ref{r1}) above.
On the other hand, the $\ket{\Phi}$ boundary
has one relevant boundary field, denoted by $\phi$,
and gives rise to a one-parameter family of reflection factors
\eq
R^{\sket{b}}(\te)= R^{\sket{\One}}(\te)/Z^{\sket{b}}(\theta)
\label{lyttrick}
\en
where the factor $Z^{\sket{b}}(\theta)$ has exactly the form
mentioned above:
\eq
Z^{\sket{b}}(\theta)=
  S(\te+i \fract{\pi}{6}(b{+}3)) S(\te- i \fract{\pi}{6}(b{+}3))
\en
where
$S(\theta)$ is the bulk S-matrix, and the parameter
$b$ can be related to the dimensionless ratio $\mu^{2}/\lambda$ of
the bulk and boundary couplings $\lambda$ and
$\mu$~\cite{Dorey:1997yg,Dorey:1999cj}.
(Since there is only one particle type in the Lee-Yang model, the
indices $a$, $d$ and so on are omitted. We also changed the
notation slightly from that of \cite{Dorey:1997yg} to avoid confusing
the parameter $b$ with a particle label.)\,\footnote{There are also
boundary-changing operators, but these were not considered in
\cite{Dorey:1997yg, Dorey:1999cj}.}

As an initial attempt to extend (\ref{lyttrick}) to
the  remaining ADET theories one
could therefore try
\eq
\R{a}{\sket{d,C}}(\te)= \R{a}{}(\te) /Z^{\sket{d,C}}_a(\theta)\,,
\label{adeSrick}
\en
with $Z_a^{\sket{d,C}}(\theta)$ as in (\ref{zone}).
This manoeuvre certainly generates mathematically-consistent sets of
reflection amplitudes, but in the more general cases it is not the
most economical choice.  Consider instead
the functions obtained by replacing
the blocks $\usa{x}$ in (\ref{smat}) by the simpler blocks $\bl{x}$
\cite{Dorey:1992bq}:
\eq
S^F_{ab}=\prod_{x\in A_{ab}}\bl{x}~.
\en
For the Lee-Yang model, $S^F(\theta)$ coincides with $S(\theta)$,
but for other theories it has fewer poles and zeros.
The bootstrap constraints are only satisfied
up to signs, but if $S^F$ is used to define a function
$Z_a^{\sket{d,C}}$ as
\eq
Z_a^{\sket{d,C}}(\theta)=
S^F_{da}(\theta+i \frac{\pi}{h} C)S^F_{da}(\theta-i
\frac{\pi}{h} C)
\label{ztwo}
\en
then these signs cancel, and thus (\ref{ztwo}) provides a more
``minimal''  generalization of the family of Lee-Yang reflection
factors which nevertheless preserves all of its desirable
properties. Setting $d$ equal to the lightest particle in the theory
generally gives the family with the smallest number of additional
poles and zeros, but we shall see evidence later that all cases have
a role to play.

The normalisation of the shift was changed in passing from
(\ref{zone}) to (\ref{ztwo}); this is convenient because,
as a consequence of the property
\eq
(x- C )(\te) \times  (x+C)(\te)=(x)(\te+ i
\fract{\pi}{h} C) \times  (x)(\te-i
\fract{\pi}{h} C)\,,
\en
we have
\eq
Z_a^{\sket{d,C}}=
\prod_{x\in A_{ad}}\bl{x- C}\bl{x+ C}\,.
\label{zdef}
\en
The factors $(Z_a^{\sket{d,C}})^{-1}$ therefore coincide with the
coupling-constant dependent parts of the affine Toda field theory
S-matrices of~\cite{Braden:1989bu,Klassen:1989ui},
with $C$ related to the parameter
$B$ of \cite{Braden:1989bu} by $C=1-B$.

\resection{Boundary $T_r$ theories as reductions of boundary sine-Gordon}
The reflection factors presented
so far are only conjectures, and no evidence has been given linking
them to any physically-realised boundary conditions. The best
signal in this respect will come from the exact $g$-function
calculations to be reported in later sections. However, for the $T_r$
theories,
alternative support for our general scheme comes
from an interesting relation with the reflection factors
of the sine-Gordon model.

In the bulk, the $T_r$ theories can be found as
particularly-simple quantum group reductions of the sine-Gordon
model at certain values of the coupling,
in which the soliton and antisoliton states
are deleted leaving only the breathers~\cite{Smirnov:1989hh}.
Quantum group reduction in the presence of
boundaries has yet to be fully understood, but the simplifications of
the $T_r$ cases allow extra progress to be made. This generalises
the analysis of
\cite{Dorey:1997yg} for $T_1$\,, the Lee-Yang model, but has some
new features.

The boundary S-matrix for the sine-Gordon solitons was found by Ghoshal
and Zamolodchikov in \cite{Ghoshal:1993tm}, and extended to the
breathers by Ghoshal
in \cite{Ghoshal:1993iq}. To match the notation used above for
the ADE theories, we trade the sine-Gordon bulk coupling
constant $\beta$ for
\eq
h=\frac{16\pi}{\beta^{2}}-2\,.
\label{hdef}
\en
Ghoshal and Zamolodchikov expressed their matrix solution to the
boundary Yang-Baxter equation for the sine-Gordon
model\footnote{also found by de Vega and Gonzalez Ruiz
\cite{deVega:1992zd}} in terms of two parameters $\xi$ and
$k$\,. However for the scalar part (which is the whole reflection
factor for the breathers) they found it more convenient to use
$\eta$ and $\vartheta$, related to $\xi$ and $k$ by
\eq
\cos\eta\,\cosh\vartheta=-\frac{1}{k}\cos\xi~,\quad
\cos^{2}\eta+\cosh^{2}\vartheta=1+\frac{1}{k^{2}}~.
\label{reparam}
\en
Ghoshal-Zamolodchikov's
reflection factor for the $a^{\rm th}$ breather on the
$\ket{\eta,\vartheta}$ boundary can then be written as
\eq
\R{a}{\sket{\eta,\vartheta}}(\theta)=
 R_0^{(a)}(\theta) R_1^{(a)}(\theta)
\label{rdef}
\en
where
\eq
R_0^{(a)}=
\frac{\bl{\frac{h}{2}}\bl{a+h}}%
{\bl{a+\frac{3h}{2}}}
\prod_{l=1}^{a-1}
\frac{\bl{l}\bl{l+h}}%
{\bl{l+\frac{3h}{2}}^{2}}
\label{r0def}
\en
is the boundary-parameter-independent part and
\eq
R_1^{(a)}(\theta)=S^{(a)}(\eta,\theta)S^{(a)}(i\vartheta,\theta)
\label{ronedef}
\en
with
\eq
S^{(a)}(\nu,\theta)=
\prod_{\substack{l=1-a \\ \textrm{step 2}}}^{a-1}
\frac{\bl{\frac{2\nu}{\pi}-\frac{h}{2}+l\,}}%
{\bl{\frac{2\nu}{\pi}+\frac{h}{2}+l\,}}
\label{ronedefb}
\en
contains the dependence on $\eta$ and $\vartheta$.
In these formulae we used the same
blocks $\bl{x}$ as defined in (\ref{bblock})
for the minimal ADE theories, with $h$ now given by (\ref{hdef}).

Now for the quantum group reduction. In the bulk, suppose that
$\beta^{2}$ is
such that
\eq
h=2r+1\,,\qquad  r\in\NN\,.
\en
At these values of the coupling, Smirnov has shown
\cite{Smirnov:1989hh} that a consistent
scattering theory can be obtained by removing all the solitonic
states, leaving $r$ breathers with masses $M_a=\frac{\sin(\pi
a/h)}{\sin(\pi/h)}\,M_1$. The scattering of these breathers is then
given by the $T_r$ S-matrix (\ref{trS}). In order to explain all poles
without recourse to the solitons,
extra three-point couplings must be introduced, some of which are
necessarily imaginary, consistent with the $T_r$ models being
perturbations of nonunitary conformal field theories.
In the presence of a boundary,
these extra couplings give rise to extra boundary bootstrap
equations, which further constrain the boundary reflection factors
and impose a relation between the parameters $\eta$ and $\vartheta$.
This can be seen by a direct analysis of the boundary
bootstrap equations but it is more interesting to take another
route, as follows.

We first recall  the observation of \cite{Dorey:1997yg}, that the
reflection factors for the boundary Lee-Yang model match solutions
of boundary Yang-Baxter equation for the sine-Gordon model at
\eq
\xi\to i\infty\,.
\label{xilim}
\en
The Lee-Yang case
corresponds to $r=1$, $h=3$, but we shall suppose that the same
constraint should hold more generally (see also
\cite{LeClair:1995uf}). The
condition must be translated into the
$(\eta,\vartheta)$ parametrisation. One degree of freedom can be
retained by
allowing $k$ to tend to infinity as the limit (\ref{xilim}) is
taken. The relations (\ref{reparam}) then become
\eq
\cos\eta\,\cosh\vartheta=A~,\quad
\cos^{2}\eta+\cosh^{2}\vartheta=1\,.
\label{qgreparam}
\en
The constant $A$ can be tuned to any value by taking $\xi$ and
$k$ to infinity suitably, and so
the first equation in (\ref{qgreparam})
is not a constraint; solving the second for
$i\vartheta$,
\eq
i\vartheta=\eta+\frac{\pi}{2}+(r{-}d)\pi~,\quad d\in\Z\,.
\label{conste}
\en
(The shift by $r$ is included for later convenience.)
This appears to give a countable infinity of one-parameter families
of reflection factors, but this is not so: the blocks $\bl{x}$ in
(\ref{ronedef}) and (\ref{ronedefb}) depend on the boundary
parameters only through the combination
\eq
\frac{2i\vartheta}{\pi}=
\frac{2\eta}{\pi}+2(r{-}d)+1\,.
\en
Since $\bl{x+2h}=\bl{x}$, the reflection factors for $d$
are therefore the same as those for $d+h$\,. In addition, the
freedom to redefine the parameter $\eta$ gives an extra invariance
of the one-parameter families under $d\to 2r-d$. Thus
the full set of options is realised by
\eq
d=0,1,\dots r\,.
\label{redop}
\en
Note also that $R^{(a)}_1(\theta)$,
the coupling-dependent part of the reflection factor,
is trivial if $d=0$. Thus the limit (\ref{xilim}) corresponds to
$r$ one-parameter families of breather reflection factors, and one
`isolated' case. This
matches the counting of conformal boundary
conditions (the set of bulk Virasoro primary fields
\cite{Cardy:1984bb}) for the ${\cal M}_{2,2r+3}$ minimal
models, and the fact that of these boundary
conditions, all but one have the relevant $\phi_{13}$ boundary
operator in their spectra.

To see how this fits in with our earlier ideas,
we
swap $\eta$ for $C$, defined by
\eq
\frac{2\eta}{\pi}=d-\frac{h}{2}+C\,.
\en
Then, starting from (\ref{ronedef}) and relabelling,
\bea
R_1^{(a)}&=&
\prod_{\substack{l=1-a \\ \textrm{step 2}}}^{a-1}
\frac{\bl{-h+d+l+C\,}}{\bl{d+l+C\,}}\,
\frac{\bl{-d+l+C\,}}{\bl{h-d+l+C\,}}\nn\\
&=&
\prod_{\substack{l=1-a \\ \textrm{step 2}}}^{a-1}
\bl{-d-l-C\,}\bl{-d-l+C\,}
\bl{-h+d+l+C\,}\bl{-h+d+l-C\,}\nn\\
&=&
\prod_{\substack{l=1-a+d\\ \textrm{step 2}}}^{a+d-1}
\bl{-l-C\,}\bl{-l+C\,}
\bl{-h+l+C\,}\bl{-h+l-C\,}\,.
\eea
Since the first $a-d$ terms in the last product cancel if
$a>d$, it is easily seen that $R_1^{(a)}$
coincides with
$(Z^{\sket{d,C}}_a)^{-1}$, as expressed by
(\ref{zdef}), for the $T_r$ S-matrix~(\ref{trS}).

\resection{Some aspects of the off-critical $g$-functions}
Our main tool in linking reflection factors to specific boundary
conditions will be the formula for an exact off-critical
$g$-function introduced in \cite{Dorey:2004xk}. In this section we
begin by reviewing the main formula, and then discuss some further
properties and modifications which will be relevant later.
\subsection{The exact $g$-function for diagonal scattering theories}
The ground-state degeneracy, or $g$-function, was introduced by
Affleck and Ludwig as a useful characterisation of the boundary
conditions in critical quantum field theories \cite{Affleck:1991tk}.
In a massive model, the definition can proceed along similar lines
\cite{LeClair:1995uf,Dorey:1997yg, Dorey:2004xk}.
One starts with the partition function
$Z_{\sbk}[L,R]$ for the theory on a
finite cylinder of circumference $L$, length $R$ and boundary
conditions of type $\alpha$ at both ends. In the $R$-channel
description, time runs along the length of the cylinder, and the
partition function is represented as a sum over an eigenbasis
$\{\ket{\psi_k}\}$ of $H^{\rm circ}(M,L)$, the Hamiltonian which
propagates states along the cylinder:
\eq
Z_{\sbk}[L,R] =   \vev{\alpha|
  \,e^{-RH^{\rm circ}(M,L)}\,
  |\alpha}  =\sum_{k=0}^{\infty}
 ~~
({\cal G}_{\sket{\alpha}}^{(k)}(l))^{2}
\, e^{ - R E_k^{\rm circ}(M,L) }\,.
\label{llchan}
\en
Here
$l=ML$, $M$ is the mass of the lightest particle in the theory, and
\eq
{\cal G}_{\sket{\alpha}}^{(k)}(l)=
\fract{\vev{\alpha|\psi_k}}{\vev{\psi_k|\psi_k}^{1/2} }~,
\en
where $\ket{\alpha}$ is the (massive) boundary state
\cite{Ghoshal:1993tm} for the $\alpha$ boundary condition.
At finite values of $l$ any possible infinite-volume vacuum
degeneracy is lifted by tunneling effects, making the ground state
$\ket{\psi_0}$ unique. This gives $Z_{\sbk}$ the following
leading and next-to-leading behaviour in the large-$R$ limit:
\eq
\ln Z_{\sbk}[L,R]
\sim - R E_0^{\rm circ}(M,L) +2\ln{\cal G}_{\sket{\alpha}}^{(0)}(l) \,.
\label{largeR}
\en
It is the second, sub-leading term which characterises the massive
$g$-function. Subtracting a linearly-growing piece $-f_{\sket{\alpha}}L$, the
$g$-function for the boundary condition $\alpha$ at system size
$l$ is defined to be
\eq
\ln g_{\sket{\alpha}}(l)= \ln {\cal G}_{\sket{\alpha}}^{(0)}(l) + f_{\sket{\alpha}} L~.
\label{gfdef}
\en
The constant  $f_{\sket{\alpha}}$ is equal to the constant
(boundary) contribution to the ground-state energy
$E^{\rm strip}_0(R)$ of the $L$-channel Hamiltonian
$H^{\rm strip}(R) \equiv H^{\rm strip}_{\sbk}(R)$, which propagates states living on a
segment of length $R$ and boundary conditions $\alpha$ at both ends.

Although this definition is in principle unambiguous, the difficulty
of characterising the finite-size states $\ket{\alpha}$ and
$\ket{\psi_0}$ limits its utility in practice.
An alternative expression for $g$ can be obtained by comparing
(\ref{largeR}) with the $L$-channel representation, a sum over the
full set $\{E^{\rm strip}_k(R)\}$ of eigenvalues of $H^{\rm
strip}(R)$\,:
\eq
 Z_{\sbk}[L,R] =
\sum_{k=0}^{\infty}\, \,
 e^{ - L E_k^{\rm strip}(M,R) }\,.
\label{rrchan}
\en
As $R$ is sent to infinity, we have
\eq
\ln g_{\sket{\alpha}}(l)= {1 \over 2}\lim_{R\rightarrow \infty} \left[
\ln \left(\sum_{k=0}^{\infty}\,
 e^{ - L E_k^{\rm strip}(M,R)} \right)+
2 f_{\sket{\alpha}} L + R E_0^{\rm circ}(M,L) \right].
\label{gadef}
\en

In theories with only massive excitations in the bulk and no
infinite-volume vacuum degeneracy,
$\ln g_{\sket{\alpha}}(l)$ tends exponentially  to zero at large $l$,
while in the UV limit $l\to 0$ it reproduces the value of the
$g$-function in the unperturbed boundary conformal field theory.

The subleading nature of the term being extracted makes it hard to
evaluate (\ref{gadef}) directly.  In \cite{Dorey:2004xk},
considerations of the infrared (large-$l$) asymptotics and a
conjectured resummation led to the proposal of a general formula for
the exact $g$-function of a purely elastic scattering theory.
However in section~\ref{secinternal} below we shall argue that the
result of \cite{Dorey:2004xk} needs to be modified by the
introduction of a simple symmetry factor whenever there is
coexistence of vacua at infinite volume. This happens in the
low-temperature phases of the
$E_6$, $E_7$,
A and D models. The more-general result is
\eq
\ln g_{\sket{\alpha}}(l) =  \ln C_{\sket{\alpha}}+\frac{1}{4}
\sum_{a=1}^{\rg}
\int_{\mathbb{R}} d\theta
 \left( \phi^{\sket{\alpha}}_{a}(\theta){-}
\delta(\theta){-}2\phi_{aa}(2\theta)\right)
     \ln \left(1+e^{-\epsilon_a(\theta)}\right)
 + \Sigma(l)
\label{exactg}
\en
where
$C_{\sket{\alpha}}$ is the symmetry factor to be discussed shortly,
and the functions
$\phi^{\sket{\alpha}}_{a}(\theta)$ and $\phi_{ab}(\theta)$ are related
to the bulk and boundary scattering amplitudes
$S_{ab}(\theta)$ and $\R{a}{\sket{\alpha}}(\theta)$ by
\bea
\phi^{\sket{\alpha}}_{a}(\theta) &=&-\frac{i}{\pi}\frac{d}{d\theta}
        \ln \R{a}{\sket{\alpha}}(\theta)\,,\\[3pt]
\phi_{ab}(\theta) &=&-\frac{i}{2\pi}\frac{d}{d\theta}\ln
S_{ab}(\theta)\,.
\eea
The boundary-condition-independent piece $\Sigma(l)$ can be expressed
in
terms of solutions to the bulk thermodynamic Bethe ansatz (TBA)
equations as
\bea
 \Sigma(l)&=& \frac{1}{2} \sum_{n=1}^{\infty}\,
   \sum_{a_1\ldots a_n=1}^{\rg} \frac{1}{n}\int_{\mathbb{R}^n}
     \frac{d\theta_1}{1+e^{\ep_{a_1}(\theta_1)}}\cdots
     \frac{d\theta_n}{1+e^{\ep_{a_n}(\theta_n)}} \times \nn \\
 & & \left(\phi_{a_1 a_2}(\theta_1+\theta_2)\phi_{a_2 a_3}(\theta_2-\theta_3)
     \ldots\phi_{a_n a_{n+1}}(\theta_n-\theta_{n+1})\right) \label{eq:gfn}
\label{exactsigma}
\eea
with $\theta_{n+1}=\theta_1$,
$a_{n+1}=a_1$\,. The
functions $\ep_a(\te)$, called pseudoenergies, solve
the bulk TBA equations
\eq
  \ep_a(\theta) =
   M_a L \cosh\te - \sum_{b=1}^{\rg} \int_{\Rth} d\te'  \,
   \phi_{ab}(\te - \te')\,
   L_b(\te')\,, ~~~~~~a=1,\dots, \rg\,,
\label{Ltbaa}
\en
where $L_a(\te)=\ln(1+e^{-\ep_a(\te)})$.
We also recall that the ground-state energy of the theory on a
circle of circumference $L$ is given in terms of these pseudoenergies
by
\eq
   E_0^{\rm circ}(M,L) = -\sum_{a=1}^{\rg}
   \int_{\Rth} {d \te \over 2\pi} \,
   M_a \cosh\theta\, L_a(\theta)
 +{\cal E}L\, M^{2}
\label{ecircl}
\en
where
${\cal E}L\, M^{2}$ is the bulk contribution to the energy.

In \cite{Dorey:2004xk} the formula (\ref{exactg}) was checked in
detail, but only for the
$\rg=1$,  $C_{\sket{\One}}=1$ case corresponding to the
Lee-Yang model. Our results below will confirm that it holds in more
general cases, provided $C_{\sket{\alpha}}$ is  appropriately chosen.

\subsection{Models with internal symmetries}
\label{secinternal}
The scaling Lee-Yang model,
on which
the analysis of \cite{Dorey:2004xk} was mostly concentrated,
is a  massive integrable quantum field
theory with fully diagonal scattering and a single vacuum.
However many models lack one or both these properties. In this
section we shall extend our analysis to models possessing, in
infinite volume, $N$ equivalent vacua  but still described by purely
elastic scattering theories\,\footnote{Integrable models with
non-equivalent vacua are typically associated to non-diagonal
scattering like, for example, the
$\phi_{13}$ perturbations of the minimal
${\cal M}_{p.q}$ models~\cite{Smirnov:1990vm}.}.
For the ADE-related theories the
$N$ equivalent vacua are related by a
global symmetry $\cal{Z}$.
We shall see that when these systems
are in their low-temperature phases, the formula for
$g(l)$  as originally proposed  in
\cite{Dorey:2004xk} should be slightly corrected,  by including
a `symmetry factor' $C_{\sket{\alpha}}$.

Low-temperature phases are common features of
two dimensional magnetic spin systems below their critical
temperatures.  A discrete
symmetry ${\cal Z}$ of the Hamiltonian $H$ is spontaneously
broken, and  a unique ground state is  singled out from a  multiplet
of equivalent degenerate  vacua. This contrasts with the
high-temperature phase, where  the ground state is  ${\cal
Z}$-invariant.

The prototype of two-dimensional  spin systems  is the Ising model,
and we shall treat this case first.
The Ising Hamiltonian for zero external magnetic field
is invariant  under a global spin reversal
transformation (${\cal Z}=\Z_2$), and at low temperatures
$T<T_c$ this symmetry is
spontaneously broken and there is a doublet
$\{
\ket{+}, \ket{-} \}$ of vacuum
states transforming into each other under a global spin-flip:
\eq
 \ket{\pm} \rightarrow  \ket{\mp}\,.
\en
The $\Z_2$ symmetry of $H$ is  preserved under renormalisation
and it also  characterizes the continuum field theory
version of the model: in the low-temperature phase
the bulk field theory has
two degenerate  vacua $\{ \ket{+}, \ket{-} \}$ with excitations, the
massive kinks $K_{[+-]}(\te)$, $K_{[-+]}(\te)$, corresponding to
field configurations interpolating between these vacua.

In infinite volume the  ground state is either
$\ket{+}$ or $\ket{-}$, and  the transition from $\ket{+}$ to
$\ket{-}$ (or vice-versa) can only happen in an  infinite interval of
time. On the other hand in a finite  volume $V$, tunneling is allowed:
a kink with finite speed  can span  the whole volume segment in a
finite time interval. We are interested in the scaling limit of the
Ising model on a finite cylinder and we interpret the open-segment
direction of length $R$ as the space coordinate. Time is then
periodic with period $L$. If the boundary conditions are taken to be
fixed, of type $+$, at both ends of the segment,
then the only multi-particle states which can propagate have
an even number of kinks, of the form
\eq
K_{[+-]}(\te_1) K_{[-+]}(\te_2) \dots  K_{[-+]}(\te_n)~,~~(\mbox{n~
even})\,.
\label{even}
\en
For $R$ large the rapidities
$\{ \te_j \}$ are quantised according to the Bethe
ansatz equations
\eq
r \sinh \te_ j -i \ln R^{\sket{+}}(\te_j) = \pi j\,,~~~(r=MR,~
j=1,2,\dots)\,,
\label{BAi}
\en
where $R^{\sket{+}}(\te)$ is the amplitude describing the
scattering of particles off  a wall with fixed boundary
conditions of type $+$\,.

At low temperatures $Z_{\sbkp}[L,R]$ therefore receives
contributions only from states with an even number
of particles. It is conveniently written in the form (see for
example \cite{Chatterjee:1995be}):
\eq
Z_{\sbkp}[L,R] = {1 \over 2} Z^{(0)}_{\sbkp}[L,R] + {1 \over 2}
Z^{(1)}_{\sbkp}[L,R]
\label{med}
\en
where
\eq
Z^{(b)}_{\sbkp}[L,R]= e^{-L E_0^{\rm strip}(M,R)} \prod_{j>0}
\left(1 +(-1)^b e^{-l \cosh \te_j} \right),~~~(l=ML)\,, \label{zz2}
\en
$E_0^{\rm strip}(M,R)$ is the ground-state energy, and the set $\{
\te_j \}$ is quantised by (\ref{BAi}). In order to  extract
the subleading contributions to
$Z_{\sbkp}$\,, as in~\cite{Dorey:2004xk} we first set
\eq
Z_{\sbkp}[L,R] = \fract{1}{2}e^{\ln Z^{(0)}_{\sbkp}[L,R]} +
\fract{1}{2} e^{\ln Z^{(1)}_{\sbkp}[L,R]}~,
\en
and in the limit $R \rightarrow \infty$ use   Newton's approximation
to transform sums into integrals. The result is
\bea
\ln Z^{(b)}_{\sbkp}[L,R] &\sim&  {1 \over 2  } \int_{\Rth} d\te
\left({r \over \pi} \cosh(\te) + \phi^{\sket{+}}(\te)  -  \delta(\te)
\right) \ln
(1 +(-1)^b e^{-l \cosh \te})\, \nn \\
&=& 2 \ln g^{(b)}_{\sket{+}}(l)  - R E_0^{{\rm
circle},b}(M,L)~,~(b=0,1)
\label{zpm}
\eea
{}From (\ref{zpm})  we see that at finite values of  $l$  as $R
\rightarrow \infty$
\eq
\ln Z^{(0)}_{\sbkp}[L,R]-\ln Z^{(1)}_{\sbkp}[L,R] \rightarrow   +\infty
\en
 and therefore
\bea
Z_{\sbkp}[L,R] &\sim& \left (g_{\sket{+}}(l) \right)^{2}  e^{-R
E_0^{\rm
circle}(M,L)}~ \sim  {1 \over 2  } Z^{(0)}_{\sbkp}[L,R] \nn \\
&=& \fract{1}{2} \Big( g_{\sket{+}}^{(0)}(l) \Big)^{2} e^{-R E_0^{\rm
circle}(M,L)}~,~
\label{f2}
\eea
where $E_0^{\rm circle}(M,L) \equiv E_0^{{\rm circle},0}(M,L)$ is the
ground state energy for the system with periodic boundary
conditions.

Notice that  $Z^{(0)}_{\sbkp}[L,R]$ takes contributions from
states with both even and odd number of particles.  In fact,
following
\cite{Dorey:2004xk}, we have $g_{\sket{+}}^{(0)}=g_{\bf
fixed}^{\hbox{ref}\cite{Dorey:2004xk}}$ and
\eq
g_{\sket{+}}(l)=g_{\sket{-}}(l)= {1 \over
\sqrt{2}} g_{\bf fixed}^{\hbox{ref}\cite{Dorey:2004xk}}~.
\label{isingc}
\en
In conclusion, the appearance of the extra symmetry factor
$C_{\sket{+}}=\frac{1}{\sqrt{2}}$ is related to the kink selection rule
restricting the possible multi-particle states. In the Ising model this
was seen in the exact formula for
the low-temperature partition function
$Z_{\sbkp}[L,R]$, by its
being written as an averaged sum of
$Z^{(0)}_{\sbkp}[L,R]$ and $Z^{(1)}_{\sbkp}[L,R]$.

In more general theories with discrete symmetries, similar
considerations apply.  Consider a set of
$i=1,2,\dots N$ `fixed' boundary conditions, matching the
$i=1,2,\dots N$ vacua. A derivation of $Z_{\sbAB{i}{i}}$
as in~\cite{Dorey:2004xk}, including {\em all}\/ multiparticle
states satisfying the Bethe momentum quantisation conditions, will
give an incorrect answer, since some states will be forbidden
by the kink structure. Instead, let us construct a new boundary
state
\eq
\ket{U} = \sum_{j=1}^N \ket{j}\,,
\en
and consider the partition function $Z_{\sbAB{U}{i}}$. Since
boundary scattering does not mix vacua it is clear that the only
effect of replacing $\ket{i}$ by $\ket{U}$ is to eliminate the kink
condition on allowed multiparticle states. Thus the counting
of states on an interval with $U$ at one end and $i$ at the other
is exactly the same as it would be in a high-temperature phase
with the reflection factor at both ends being
$R_{\sket{i}}(\te)$, and the derivation
in~\cite{Dorey:2004xk} goes through to find at large $R$
\eq
\ln Z_{\sbAB{U}{i}}[L,R]
\sim  - R E_0^{\rm circ}(M,L) +2\ln{\cal G}^{(0)}(l) \,
\en
where ${\cal G}^{(0)}$ is -- up to a linear term -- given
by~(\ref{exactg}) with
$\phi^{\sket{\alpha}}_{a}(\te)=\phi^{\sket{i}}_{a}(\te)$ and
$C_{\sket{\alpha}}=1$. On the other
hand, since all vacua are related by the discrete symmetry $\cal Z$
and the finite-volume vacuum state
$\ket{\psi_0}$ must be symmetrical under
this symmetry,
$\vev{\psi_0|i} =
\vev{\psi_0|j}$ $\forall\, {i,j}$, and so
\eq
\left({\cal G}^{(0)}(l) \right)^{2}={\vev{U|\psi_0} \vev{\psi_0|i} \over
\vev{\psi_0|\psi_0}}= N {\vev{\psi_0|i}^{2} \over
\vev{\psi_0|\psi_0}}= N \left ({\cal G}_{\sket{i}}^{(0)}(l)
\right)^{2}.
\en
Hence
\eq
g_{\sket{i}}(l) =  \fract{1}{\sqrt{N}} g(l)~
\en
and $g_{\sket{i}}(l)$ is given by the formula
(\ref{exactg}) with
$C_{\sket{i}}=\fract{1}{\sqrt{N}}$.

An immediate  consequence is that in the infrared,
$g_{\sket{i}}(l)$ does not tend to one. Instead,
\eq
\lim_{l \rightarrow \infty} g_{\sket{i}}(l) = C_{\sket{i}}={1 \over
\sqrt{N}}\,.
\label{asymg}
\en
This perhaps-surprising claim can be justified independently.
In infinite volume, boundary states can be described using a basis
of scattering states \cite{Ghoshal:1993tm}:
\eq
\ket{B}=\left(1+\frac{1}{2}\int^{\infty}_{-\infty}
K^{ab}(\theta)A_a(-\theta)A_b(\theta)+\dots\right)\ket{0}
\label{bstate}
\en
where $\ket{0}$ is the bulk vacuum for an infinite line.
In finite but large volumes,
the same expression provides a good approximation to the boundary
state, the only modification normally being that the momenta of the
multiparticle states must be quantised by the relevant Bethe ansatz
equations.
However a subtlety
arises when there is a degeneracy among the bulk vacua, in situations
where the boundary condition distinguishes between these vacua.
(Similar issues are encountered in comparing exact bulk VEVs of
fields with those obtained from finite-volume approximations
\cite{Fateev:1997yg,Guida:1997fs}.)
Take
the case of a fixed boundary condition which picks out one of the
degenerate vacua, say $i$: then the state $\ket{0}$ on the RHS of
(\ref{bstate}) is $\ket{i}$. However when calculating the
finite-volume $g$-function, we must consider $\vev{\psi_0|B}$, where
$\bra{\psi_0}$
is the finite-volume vacuum.
(Note that the bulk energy is normalised to zero when considering
(\ref{bstate}), so this inner product will give us $g$ directly,
rather than $\cal G$.)
In large but finite volumes, the tunneling amplitude between the
infinite-volume bulk
vacua is non-zero and so the appropriately-normalised finite volume
ground state is the symmetric combination
\eq
\bra{\psi_0}=\frac{1}{\sqrt{N}}\sum_{j=1}^N\bra{j}\,.
\en
The limiting value of $g$ in the infrared is therefore not $1$,
but $1/\sqrt{N}$, as found in the exact calculation
earlier.

Table~\ref{tabb1}  lists  the central charge, the Coxeter number, the
symmetry group
$\cal Z$, and the number of degenerate vacua for the ADET models.
For the $A_r$, $D_r$,
$E_6$ and $E_7$ theories at low temperatures there is  a coexistence of
$N$ vacuum states
$\ket{i}$, $i=1,2,\dots,N$,
with $N=r+1$, $4$, $3$
and $2$ respectively. The corresponding symmetry factor
$C_{\sket{i}}$\,, for fixed-type boundary conditions which single out
a single bulk vacuum, is always $1/\sqrt{N}$. (For more general
boundary conditions one can anticipate other symmetry factors, but we
shall leave a detailed discussion of this point for another time.)

\begin{table}[tb]
\begin{center}
\begin{tabular}{|c|c|c|c|c|}
\hline
\rll
\small Model &\small $c_{\rm eff}(0)$ &\small  Coxeter number $h$
&\small  Global symmetry $\cal Z$ &
\small {}~~~~$N${}~~~~\\
\hline
\hline
\rl
 $A_r$ & $\frac{2r}{r+3}$  & $r+1$  & $\Z_{r+1}$ & $r+1$ \\
\hline
\rl
 $D_r$~($r$ even)  &  $1$ & $2r-2$ & $\Z_2 \times \Z_2$ & 4 \\
\hline
\rl
 $D_r$~($r$ odd)   &  $1$ & $2r-2$ & $\Z_4$ & 4 \\
\hline
\rl
 $E_6$  &  $\frac{6}{7}$ & 12  & $\Z_3 $ & 3  \\
\hline
\rl
 $E_7$  & $\frac{7}{10}$   & 18 & $\Z_2 $ & 2 \\
\hline
\rl
 $E_8$  & $\frac{1}{2}$   & 30 & $\Z_1 $ & 1\\
\hline
\rl
 $T_r$  &  $\frac{2r}{2r+3}$  & $2r+1$ & $\Z_1 $ & 1 \\
\hline
\end{tabular}
\caption{\small Data for the ADET purely elastic scattering theories.
The symmetry factor $C_{\sket{i}}$ for fixed-type boundary conditions
is $1/\sqrt{N}$ in each case, where $N$ is the number of degenerate
vacua in the low-temperature phase. }
\label{tabb1}
\end{center}
\end{table}
\subsection{Further properties of the exact $g$-function}
\label{sprope}
An important property of the sets of TBA equations that we are
considering concerns the so-called Y-functions
\cite{Zamolodchikov:1991et},
\eq
Y_a(\te)=e^{\ep_a(\te)}.
\label{Ye}
\en
These satisfy a set of functional relations called a
Y-system:
\eq
Y_a(\te+{i \pi \over h})Y_a(\te-{i \pi \over h})=\prod_{b=1}^{\rg}
\left(1+ Y_b(\te)
\right)^{A^{\bf[G]}_{ba}}~,
\en
where $A^{\bf [G]}_{ba}$ is the incidence matrix of the Dynkin
diagram
$G$ labelling the TBA system.
Defining an associated set of T-functions through the relations
\eq
Y_a(\te)= \prod_{c=1}^{\rg} \left( T_c(\te)
\right)^{A^{\bf [G]}_{ca}}~~,~~1+Y_a(\te)=
T_a(\te+i \fract{\pi}{h}) T_a(\te-i \fract{\pi}{h})
\label{YT}
\en
we have
\eq
T_a(\te+i \fract{\pi}{h})T_a(\te-i \fract{\pi}{h})
\prod_{c=1}^{\rg}\left( T_c(\te)
\right)^{-A^{\bf [G]}_{ca}}=1+Y_a^{-1}(\te)~.
\label{ty}
\en
In addition the  T-functions satisfy
\eq
T_a(\te+\frac{i \pi}{h}) T_a(\te-\frac{i\pi}{h})=1+
\prod_{b=1}^{\rg} \left(T_b(\te)
\right)^{A^{\bf [G]}_{ba}}~.
\en
Fourier transforming the logarithm
of equation (\ref{ty}), solving taking the
large $\theta$ asymptotic into account, and transforming back
recovers the (standard) formula
\eq
\ln T_a(\te)= { L M_a \over 2 \cos \fract{\pi}{h}} \cosh \te -
\sum_{b=1}^{\rg} \int_{\Rth}
d \te'  \,
   \chi_{ab}(\te - \te')\,
   L_b(\te')\,, ~~~~~~a=1,\dots\rg~,~~~
\label{tint}
\en
where~\cite{Zamolodchikov:1991et,Ravanini:1992fi}
\eq
\chi_{ab}(\te)=-\int_{\Rth}{dk \over 2 \pi}
\,e^{ik\theta}\left(2\cosh(k\pi/h)\One -A^{\bf [G]}\right)^{-1}_{ab}
= - \frac{i}{2\pi}\frac{d}{d\te} S^F_{ab}(\te)\,.
\en
The result (\ref{tint}) allows us to make a connection between the
exact
$g$-functions for our parameter-dependent reflection factors and the T
functions. Taking the
reflection factors defined by (\ref{ztwo}) and (\ref{adeSrick})
\eq
R^{\sket{b,C}}_a=R^{(a)}(\te)/(S^F_{ab}(\te-i
\fract{\pi}{h} C)S^F_{ab}(\te+i\fract{\pi}{h} C))
\en
and using~(\ref{exactg}), we find
\bea
\ln g_{\sket{b,C}}(l)&=& \ln g(l) -
\frac{1}{2}\sum_{a=1}^{\rg}\int_{\Rth} d\te'
\chi_{ab}(\te'-i \fract{\pi}{h} C) L_b(\te') \nn\\
&&\qquad\qquad {~} -  \frac{1}{2} \sum_{a=1}^{\rg}\
\int_{\Rth} d\te' \chi_{ab}(\te'+i\fract{\pi}{h} C) L_b(\te')\,.
\eea
Comparing this result with  (\ref{tint}) and using the property
$T_b(\te)=T_b(-\te)$,
\eq
\ln g_{\sket{b,C}}(l) =\ln g(l)+
\ln T_{b}(i\frac{\pi}{h} C)-\frac{ L M_b }{ 2 \cos \fract{\pi}{h}}
\cos(\frac{\pi}{h}  C)
\en
or, subtracting the linear contribution:
\eq
{\cal G}_{\sket{b,C}}^{(0)}(l)={\cal G}^{(0)}(l) T_b(i\frac{\pi}{h}
C)\,.
\en
Exact relations between $g$- and T- functions in various situations
where the bulk remains critical were observed in
\cite{Fendley:1994rh,Bazhanov:1994ft,Bazhanov:2001xm}. These were
extended off-criticality to a relation between the ${\cal G}$- and
$T$- function of the Lee-Yang model
in~\cite{Dorey:1999cj}. The generalisation of \cite{Dorey:1999cj}
proposed here relies on the specific forms of our one-parameter
families of reflection factors, and thus provides some further motivation
for their introduction.

It is sometimes helpful to have an alternative representation for
the infinite sum $\Sigma(l)$ in (\ref{exactg}), the piece of the
$g$-function which does not depend on the specific boundary condition.
It is readily checked that, for values of $l$ such that the sum
converges,
\eq
\Sigma(l)= \sum_{n=1}^{\infty}\frac{1}{n} \mbox{Tr} K^n -
\sum_{n=1}^{\infty} \frac{1}{2n} \mbox{Tr} H^n
\en
where
\eq
K_{ab}(\te,\te')=  \frac{1}{2}{1 \over \sqrt{1+ e^{\ep_a(\te)}}}
\Big( \phi_{ab}(\te+\te') + \phi_{ab}(\te-\te')
\Big)\frac{1}{\sqrt{1+ e^{\ep_b(\te')}}}
\en
and
\eq
H_{ab}(\te,\te')=  \frac{1}{\sqrt{1+ e^{\ep_a(\te)}}}
\phi_{ab}(\te-\te')\frac{1}{\sqrt{ 1+ e^{\ep_b(\te')}}}\,.
\en
The identity
\eq
\ln \mbox{Det}(I-M)= \mbox{Tr} \ln(I-M) = - \sum_{n=1}^{\infty}
\frac{1}{n} \mbox{Tr} M^n
\en
then allows $\Sigma(l)$ to be rewritten as
\eq
 \Sigma(l)=\frac{1}{2}\ \ln \mbox{Det} \Big({I-H \over (I-K)^{2}} \Big)\,.
\label{exactgdet}
\en
This formula makes sense even when the original sum
diverges, a fact that will be used in the next section.

\resection{UV values of the $g$-function}
Taking into account the effect of vacuum degeneracies described above,
a first test of
the reflection factors found earlier is to calculate the
$l\rightarrow 0$ limit of (\ref{exactg}).
The value of
$g_{\sket{\alpha}}(0)$ should match the value of a conformal field
theory $g$-function, either of a Cardy state or of a superposition of
such states.

As $l\to 0$, the pseudoenergies
$\ep_a(\theta)$ tend to constants, which we shall denote as
$\epsilon_a$.  Their values
were tabulated by Klassen and Melzer in~\cite{Klassen:1989ui},
who also observed an elegant formula the integrals of the logarithmic
derivatives of the bulk S-matrix elements, later proved in
\cite{Dorey:1991zp}\,:
\eq
\int_{\mathbb{R}} \,d\theta\,\phi_{ab}(\theta)
=-{\cal N}_{ab}\,,
\label{kmformula}
\en
where ${\cal N}_{ab}$ is related to the Cartan matrix ${\cal C}$ of
the associated  Lie algebra by
${\cal N} =2{\cal C}^{-1}-1$.\,\footnote{The Cartan matrix for $T_r$ is
taken to be
$\left(\begin{array}{ccccc}
2 & -1 & & & \\
-1 & 2 & -1 & &  \\
 & &\ddots & & \\
 & & & 2 & -1 \\
 & & & -1 & 1 \\
\end{array} \right)$ }
We shall also need the integrals of the logarithmic derivatives of the
reflection factors. Writing
\eq
R_a = \prod_{x\in A} (x)
\en
for some set $A$, the required integrals are
\eq
\int_{\mathbb{R}} d\theta\,\phi^{\sket{\alpha}}_a(\theta)
 =-2\sum_{x\in A}\left (1-\frac{x}{h} \right)\; \hbox{sign}[x/h]
\en
with $\hbox{sign}[0]=0$. For the minimal
ADE reflection factors found in section
\ref{sec.R} these evaluate to
\eq
\int_{\mathbb{R}} d\theta\,\phi^{\sket{\alpha}}_a(\theta)
=1-{\cal N}_{aa}\,,
\label{aderefl}
\en
while for the $T_r$ reflection factors (\ref{r0def}),
\eq
\int_{\mathbb{R}} d\theta\,\phi^{\sket{\alpha}}_a(\theta)
=-2{\cal N}_{aa}\,.
\en
To calculate the UV limit of $\Sigma(l)$,
we define a matrix ${\cal M}$ whose elements are
\eq
{\cal M}_{ab}=-\frac{{\cal N}_{ab}}{1+e^{\epsilon_a}}\,,
\en
in terms of which
\eq
\Sigma(0)=
\frac{1}{4}\sum_{n=1}^\infty \frac{1}{n}(e_1^n+\ldots+e_r^n)
=-\frac{1}{4} \ln\left((1-e_1)\ldots(1-e_r)\right)
\label{sumin}
\en
where $e_1,\ldots,e_r$ are the eigenvalues of ${\cal M}$. When one
of these eigenvalues is larger than 1, the infinite sum does not
converge (this is the case for $A_r$,
$r\geq 5$, $D_r$,
$r\geq 4$ and $E_r$)
but  the RHS of (\ref{sumin}) gives the correct
analytic continuation, since it follows from
(\ref{exactgdet}).

\subsection{$T_r$}
We start with the $T_r$ theories, whose UV limits are
the non-unitary minimal models ${\cal M}_{2,2r+3}$.
There are $r+1$ `pure' conformal boundary conditions,
and the corresponding values of the conformal $g$-functions can
be found using the formula
\eq
g_{(1,1+d)}=\frac{S_{(1,1+r);(1,1+d)}}{|S_{(1,1+r),(1,1)}|^{1/2}}\quad
,\quad d=0,\ldots,r
\label{gmin}
\en
where $(1,1+r)$ is the ground state of the bulk theory, with
lowest conformal weight,
$(1,1)$ is the `conformal vacuum' -- with
conformal weight $0$ -- and $S$ is the modular S-matrix
(see for example \cite{DiFrancesco:1997nk}).
The components of $S$ for a general minimal model ${\cal M}_{p'p}$
are
\eq \label{Smin}
S_{(n,m);(\rho,\sigma)}=
2\sqrt{\frac{2}{pp'}}(-1)^{1+m\rho+n\sigma}
\sin(\pi\frac{p}{p'}n\rho)\sin(\pi\frac{p'}{p}m\sigma)~,
\en
where $1\leq n,\rho\leq p'-1$, $1\leq m,\sigma\leq p-1$ and, to avoid
double-counting, $m<\frac{p}{p'}n$ and $\rho<\frac{p}{p'}\sigma$.
For the $T_r$ theories,
$p'=2$ and $p=2r+3$, and so $n$ and $\rho$ are both equal to $1$
while $m$ and $\sigma$ range from $1$ to $r+1$, as in (\ref{gmin}).

The values predicted by (\ref{gmin}) should be compared with
the values of $g_{\sket{\alpha}}(0)$ calculated from
(\ref{exactg}). For the
boundary-parameter-independent part of the reflection factor from
(\ref{r0def}), which is the $d=0$ case of the
`quantum group reduced' options (\ref{redop}), the second term
of (\ref{exactg}) evaluates to
\bea
& &\frac{1}{4}\sum_{a=1}^r\int_{\mathbb{R}}d\theta
 \left( \phi^{\sket{\alpha}}_a(\theta)-\delta(\theta)
 -2\phi_{aa}(2\theta)\right)
 \ln\left(1+e^{-\epsilon_a}\right) \nn \\
& &\qquad\qquad  =-\frac{1}{2}\sum_{a=1}^r a\ln\left(1+
 \frac{\sin^{2}\left(
 \frac{\pi}{2r+3}\right)}{\sin\left(\frac{a\pi}{2r+3}\right)\sin\left(
 \frac{(a+2)\pi}{2r+3}\right)}\right) \nn \\
& &\qquad\qquad = \ln \left(\frac{\sin\left(\frac{\pi}{2r+3}\right)
 \sin^r\left(\frac{(r+2)\pi}{2r+3}\right)}
 {\sin^{r+1}\left(\frac{(r+1)\pi}{2r+3}\right)}\right),
\eea
while $\Sigma(0)$,
calculated using (\ref{sumin}), is
\eq
-\frac{1}{4}\ln\left((1-e_1)\ldots(1-e_r)\right)= \frac{1}{2}
\ln\left(\frac{2}{\sqrt{2r{+}3}}\sin\left(\frac{(r{+}1)\pi}
{2r{+}3}\right)\right)~.
\en
Adding these
terms together and using some simple trigonometric identities reveals
a dramatic simplification:
\eq
g(l)|_{l=0} = \left(\frac{2}{\sqrt{2r{+}3}}
\sin\left(\frac{\pi}{2r{+}3}\right)
\right)^{\frac{1}{2}}
= g_{(1,1)}
\en
for all $r$. This suggests that the minimal $T_r$ reflection
factors (\ref{ronedef}) describe bulk perturbations of the boundary
conformal field theory with $(1,1)$ boundary conditions. Notice that
among all of the possible conformal boundary conditions in the
unperturbed theory, this is the
only one with no relevant boundary operators, matching the fact that
the
minimal reflection factors (\ref{ronedef}) have no free parameters.

\subsection{$A_r$, $D_r$ and $E_r$}
The ADE cases exhibit an interestingly uniform structure:
substituting (\ref{kmformula}) and (\ref{aderefl}) into (\ref{exactg})
shows that the UV limit of
the second, reflection-factor-dependent, term of (\ref{exactg}) is
always zero when the minimal reflection
factors from section~\ref{sec.R} are used.
This result can be confirmed by examining the contributions of
the blocks $\bl{x}$ of $S(2\theta)$, and $\blk{x}$ (or $\tbl{x}$) of
$R^{\sket{\alpha}}(\theta)$, as follows.
Since the bulk S-matrix can be written as
\eq
S_{ab}(\theta)=\prod_{x\in A_{ab}}\ubl{x}(\te)
\en
where $\ubl{x}=\bl{x+1}\bl{x-1}$,
$S_{ab}(2\theta)$ can be written similarly:
\eq
S_{ab}(2\theta)=\prod_{x\in A_{ab}}[x](\te)
\en
where
\eq
[x]=\left(\frac{x+1}{2}\right)\left(\frac{x-1}{2}\right)
\left(\frac{2h-x+1}{2}\right)^{-1}\left(\frac{2h-x-1}{2}\right)^{-1}~.
\en
Summing the contributions of each block $[x]$, it
is easily seen that
\eq
2 \int_{\mathbb{R}} d\theta\,\phi_{aa}(2\theta) = \sum_{x
\in A_{aa}}\left(\frac{2x}{h}
+\delta_{x,1}-2\right).
\en
The minimal reflection factors discussed in section \ref{sec.R} can be
written in a similar way:
\eq
R_{a}=\prod_{x\in A_{aa}}f_x
\en
where each $f_x$ is either $\blk{x}$ or $\tbl{x}$. The
contribution to $\int d\theta\,\phi^{\sket{\alpha}}_a(\theta)$ from each
$\blk{x}$ is $-2+2x/h+2\delta_{x,1}$ and from each $\tbl{x}$ is $-2+2x/h$.
Noting that every minimal reflection factor contains the block
$\blk{1}$ exactly once, we find
\eq
\int_{\mathbb{R}} \, d\theta\,\phi^{\sket{\alpha}}_a(\theta)
 =\sum_{x\in A_{aa}}\left(\frac{2x}{h}
+2\delta_{x,1}-2\right),
\en
and so
\eq
\int_{\mathbb{R}}d\theta
\left( \phi^{\sket{\alpha}}_a(\theta)-\delta(\theta)-2\phi_{aa}(2\theta)\right)
=0
\en
for every A, D and E theory, as claimed. Working backwards, this
gives a general proof of the formula
(\ref{aderefl}), given
(\ref{kmformula}).

With the reflection-factor-dependent term giving zero,
the sum of $\Sigma(0)$ and the
symmetry factor should correspond to a conformal $g$-function
value.
To find these values in a uniform way, we shall use
the diagonal coset description $\hat{\mathfrak{g}}_1\times
\hat{\mathfrak{g}}_1/\hat{\mathfrak{g}}_2$ where $\hat{\mathfrak{g}}_l$ is the
affine Lie algebra at level $l$ associated to one
of the A, D or E Lie algebras \cite{Goddard:1984vk}.
Coset fields are specified by triples $\{\bmu,\bnu;\brho\}$
of $\hat{\mathfrak{g}}$ weights at levels $1$, $1$ and $2$, and
the $g$-function for the corresponding conformal boundary condition
can be written in terms of the modular S-matrix of
the coset model
$S_{\{\bzero,\bzero;\bzero\}\{\bmu,\bnu;\brho\}}$ as
\eq \label{gcoset}
g_{\{\bmu,\bnu;\brho\}}=
\frac{S_{\{\bzero,\bzero;\bzero\}\{\bmu,\bnu;\brho\}}}
{\sqrt{S_{\{\bzero,\bzero;\bzero\}\{\bzero,\bzero;\bzero\}}}}\,.
\en
The integrable highest weights $\bmu$ of
$\hat{\mathfrak{g}}$ can be expressed in terms of the fundamental weights
$\bLambda_i$ of $\hat{\mathfrak{g}}$ as:
\eq
\bmu = \sum_{i=0}^{\rg} n_i\bLambda_i
\en
where the non-negative integers
$n_i$, $i=0,\ldots,{\rg}$, are the Dynkin labels. The
fundamental weights can be written as
\bea
\bLambda_0 &=& (\bzero,1,0) \\
\bLambda_i &=& (\blambda_i,\check{a}_i,0) \nn
\eea
where $\blambda_i$ are the fundamental weights of $\mathfrak{g}$ and
the colabels, $\check{a}_i$, are given for each case
below:
\begin{align}
A_r\, :\, & \check{a}_i = 1 \quad\textrm{for all $i$} \nn \\
D_r\, :\, & \check{a}_1 = \check{a}_{r-1} = \check{a}_r = 1
    \quad\textrm{all others are 2} \nn \\
E_6\, :\, & \check{a}_i = (1,1,2,2,2,3) \\
E_7\, :\, & \check{a}_i = (1,2,2,2,3,3,4) \nn \\
E_8\, :\, & \check{a}_i = (2,2,3,3,4,4,5,6).\nn
\end{align}
The highest weights $\bmu_0$ of the corresponding finite-dimensional
Lie algebra, $\mathfrak{g}$, can be written
in a similar way, with Dynkin labels, $n_1,\ldots,n_{\rg}$, taken
from the same set:
\eq
\bmu_0 = \sum_{i=1}^{{\rg}}n_i\blambda_i.
\en
Once the level $l$ of $\hat{\mathfrak{g}}$ has been specified,
the Dynkin labels must satisfy
\eq
l = n_0 +\sum_{i=1}^{{\rg}}n_i \check{a}_i
\en
so the representation of
$\hat{\mathfrak{g}}_l$ is completely determined by the Dynkin
labels of the corresponding representation of $\mathfrak{g}$.
For example, the label $\bzero$ is given to the vacuum representation,
where $n_i=0$, $i=1,\ldots,{\rg}$ for both $\hat{\mathfrak{g}}_1$
and $\hat{\mathfrak{g}}_2$.

In appendix~\ref{coset} we show among other things
 that the coset conformal $g$-function~(\ref{gcoset}) depends only
on the level 2 weight, via the level 2 modular S-matrix\footnote{We'd
like to thank Daniel Roggenkamp for his help in explaining this
observation.}:
\eq\label{g}
g_{\{\bmu,\bnu;\brho\}} =
g_{\brho} =
\frac{S^{(2)}_{\bzero\brho}}{\sqrt{S^{(2)}_{\bzero\bzero}}}.
\en
Clearly,
this will not fix the coset field in general. For example, the number
of boundary conditions with $g$-function equal to $g_{\bzero}$
is $r+1$, 4, 3, 2, 2 for the $A_r$, $D_r$,
$E_6$, $E_7$ and $E_8$ models respectively. More details on
this point are given in
appendix~\ref{coset}.

Algorithms for computing the modular S-matrices are given by Gannon
in~\cite{Gannon:2001py};
Schellekens has also produced a useful program for their
calculation \cite{Schell}. The level 2  modular
S-matrix elements for A,
D and E theories needed to calculate the UV values of the
$g$-functions using eq.(\ref{g}) are given in table~\ref{tabb2}. The
representations are labelled by the Dynkin labels, $n_i$, $i=0,\ldots,r$.
The number of coset fields corresponding to each label is equal to the
order of the orbit of that level 2 weight, under the outer automorphism group
$O(\hat{\mathfrak{g}}$).
Note for $D_r$, $r$ even, the weights $\brho_{r-1}$ and $\brho_r$ are in
different orbits, each with order 2, whereas for $r$ odd they are in
the same orbit with order 4, so in both cases there are 4 coset fields with
the same $g$-function value.

\begin{table}[tbp]
\small
\hskip -10pt
\begin{tabular}{|c|c|c|}
\hline
 &\rl S-matrix element & Labels \\[-7pt]
\hline
\hline
$A_r$ &
 $S^{(2)}_{\bzero\bzero} = \frac{2^{\frac{h+3}{2}}}{(h+2)\sqrt{h}}
 \sin\left(\frac{\pi}{h+2}\right)
 \prod_{k=1}^{\left[\frac{h+1}{2}\right]}\sin\left(\frac{k\pi}{h+2}\right)$
 & $\bzero\,:\,n_0=2\,,\, n_i=0$ for $i=1,\ldots,r$ \\
 & &
$\brho_j\,:\,n_0=1\,,\, n_i = \left\{\begin{array}{ll}
      0 & i\neq j \\
      1 & i=j
       \end{array} \right.$ \\
 &$S^{(2)}_{\bzero\brho_j} = \frac{2^{\frac{h+3}{2}}}{(h+2)\sqrt{h}}
 \sin\left(\frac{(j+1)\pi}{h+2}\right)
 \prod_{k=1}^{\left[\frac{h+1}{2}\right]}\sin\left(\frac{k\pi}{h+2}\right)$
 & for $j=1,\ldots,[h/2]$\,, \\
 & & where $[x]$ is the integer part of $x$ \\
 & & and $h$ is the Coxeter number. \\
\hline
$D_r$ & $S^{(2)}_{\bzero\bzero} = \frac{1}{4}\sqrt{\frac{2}{r}}$ &
$\bzero\,:\,n_0=2\,,\,n_i=0$ for $i=1,\ldots,r$ \\
& &$\brho_1\,:\, n_0=n_1 =1\,,\,\textrm{all other $n_i=0$}$ \\
 & $S^{(2)}_{\bzero\brho_1}=S^{(2)}_{\bzero\brho_j} = 2S^{(2)}_{\bzero\bzero}$
 &   $\brho_j\,:\,n_0=0\,,\,n_i = \left\{\begin{array}{ll}
      0 & i\neq j \\
      1 & i=j
       \end{array} \right.$ \\
 & & for $j=2,\ldots,r/2$ \\
 & $S^{(2)}_{\bzero\brho_{r-1}}=S^{(2)}_{\bzero\brho_{r}} = \frac{1}{2\sqrt{2}}$ &
 $\brho_{r-1}:\,n_0=n_{r-1} =1\,,\,\textrm{all other $n_i=0$}$\\
 & & $\brho_{r}\,:\,n_0=n_r =1\,,\,\textrm{all other $n_i=0$}$ \\
\hline
$E_6$ & $S^{(2)}_{\bzero\bzero} = \frac{2}{\sqrt{21}}
\sin\left(\frac{2\pi}{h+2}\right)$ &
$\bzero:\,n_0=2\,,\,n_i=0$ for $i=1,\bar{1},2,3,\bar{3},4$ \\
 & $S^{(2)}_{\bzero\brho_j} = \frac{2}{\sqrt{21}}
\sin\left(\frac{2(4-j)\pi}{h+2}\right)$ for $j=1,2$ &
$\brho_1\,:\,n_0=1\,,\,n_1=1$ all other $n_i=0$ \\
 & &
$\brho_2\,:\,n_0=0\,,\,n_2=1$ all other $n_i=0$ \\
\hline
$E_7$ &
$S^{(2)}_{\bzero\bzero} = \frac{2}{\sqrt{h+2}}\sin\left(\frac{4\pi}{h+2}\right)$ &
$\bzero\,:\,n_0=2\,,\,n_i=0$ for $i=1,\ldots,7$ \\
 & $S^{(2)}_{\bzero\brho_1} = 2\sqrt{\frac{2}{h+2}}
\sin\left(\frac{8\pi}{h+2}\right)$ &
$\brho_1\,:\,n_0=1\,,\,n_1=1$ all other $n_i=0$ \\
 & $S^{(2)}_{\bzero\brho_2} = \frac{2}{\sqrt{h+2}}
\sin\left(\frac{8\pi}{h+2}\right)$ &
$\brho_2\,:\,n_0=0\,,\,n_2=1$ all other $n_i=0$ \\
 & $S^{(2)}_{\bzero\brho_3} = 2\sqrt{\frac{2}{h+2}}
\sin\left(\frac{4\pi}{h+2}\right)$ &
$\brho_3\,:\,n_0=0\,,\,n_3=1$ all other $n_i=0$ \\
\hline
$E_8$ & $S^{(2)}_{\bzero\bzero}=S^{(2)}_{\bzero\brho_2} = \frac{1}{2}$ &
$\bzero:\,n_0=2\,,\,n_i =0\,,\,\textrm{for all $i=1,\ldots,8$}$ \\
 & & $\brho_2\,:\,n_0=0\,,\,n_2 =1\,,\,\textrm{all other $n_i=0$}$ \\
 & $S^{(2)}_{\bzero\brho_1} = \frac{1}{\sqrt{2}}$ &
$\brho_1\,:\,n_0=0\,,\,n_1=1$ all other $n_i=0$ \\
\hline
\end{tabular}
\caption{\small Level 2 modular S-matrix elements for A, D and E models}
\label{tabb2}
\end{table}

For each coset, the possible
CFT values of the $g$-function can now
be calculated using eq.(\ref{g}) and the modular S-matrix
elements given in table~\ref{tabb2}.
Working case-by-case, we checked these numbers against
the sums $\ln C_{\sket{\alpha}}+\Sigma(0)$,
the UV limits of the
off-critical $g$-functions $g(l)$
for the minimal reflection factors
described in section~\ref{sec.R}.
In every case, we found that
\eq
\ln g(l)|_{l=0}=
\ln C_{\sket{\alpha}}+\Sigma(0)=
\ln  g_{\bzero}
\en
provided that the symmetry factors $C_{\sket{\alpha}}$ were assigned
as in table
\ref{tabb1}.
The explicit $g$-function values are given in table~\ref{tabb3}.

\begin{table}[htbp]
\begin{center}
\begin{tabular}{|c|c|c|c|}
\hline
 &\rl\small $C_{\sket{\alpha}}$ &\small $g_{\bzero}$
&\small ~Number of fields~ \\[-5pt]
\hline
\hline
$A_r$ & $\frac{1}{\sqrt{r+1}}$ & $\left(\frac{2^{\frac{h+3}{2}}}{(h+2)\sqrt{h}}
 \sin\left(\frac{\pi}{h+2}\right)
 \prod_{k=1}^{\left[\frac{h+1}{2}\right]}\sin\left(\frac{k\pi}{h+2}\right)
 \right)^{1/2}$ & $r+1$ \\
\hline
$D_r$ & $\frac{1}{2}$ & $\frac{1}{2}\left(\frac{2}{r}\right)^{1/4}$ & 4 \\
\hline
$E_6$ & $\frac{1}{\sqrt{3}}$ & $\left(\frac{2}{\sqrt{21}}\sin
\left(\frac{2\pi}{h+2}\right)\right)^{1/2}$ & 3 \\
\hline
$E_7$ & $\frac{1}{\sqrt{2}}$ & $\left(\frac{2}{\sqrt{h+2}}\sin
\left(\frac{4\pi}{h+2}\right)\right)^{1/2}$ & 2 \\
\hline
$E_8$ & 1 & $\frac{1}{\sqrt{2}}$ & 2 \\
\hline
\end{tabular}
\caption{\small
UV $g$-function values calculated from the minimal reflection factors
for the A, D and E models}
\label{tabb3}
\end{center}
\end{table}

For some minimal models,
we can compare $g_{\bzero}$ to the values of
the $g$-function corresponding to known cases, thereby
matching our reflection factors to physical boundary conditions. For the
three-state Potts model ($A_2$), the tricritical Ising model ($E_7$)
and the Ising model ($E_8$) the corresponding boundary condition is
the `fixed' condition in each
case~\cite{Affleck:1998nq,Chim:1995kf,Affleck:1991tk}. Note the number
of coset fields with $g$-function equal to $g_{\bzero}$
corresponds to the number of
degenerate vacua, and hence to the number of possible `fixed' boundary
conditions for all $A$, $D$ and $E$ models.

\resection{Checks in conformal perturbation theory}
The off-critical $g$-functions only match boundary
conformal field theory values in the far ultraviolet. Moving away
from this point one expects a variety of corrections,
some of which were analysed
using conformal perturbation theory in \cite{Dorey:1999cj}.
The expansion provided by our exact $g$-function result is instead
about the infrared, but convergence is sufficiently fast that
the first few terms of the UV expansion can be extracted numerically,
allowing a
comparison with conformal perturbation theory to be made. In
\cite{Dorey:2004xk} this was done for the boundary Lee-Yang model; in
this section we test our more general proposals for the case of the
three-state Potts model. Since the treatment of conformal
perturbation theory in \cite{Dorey:1999cj} concentrated on the
non-unitary Lee-Yang case, we begin with a general
discussion of the leading bulk-induced
correction to the $g$-function in the
(simpler to treat) unitary cases.

Consider a unitary conformal field theory
on a circle of circumference $L$,
perturbed by a bulk spinless primary field $\varphi$ with scaling
dimension
$x_{\varphi}=\Delta_{\varphi}+\overline\Delta_{\varphi}$\,.
The perturbed Hamiltonian is then
\eq
\hat H=\hat H_0 + \lambda\,\hat H_1
\en
where
\eq
\hat H_0=\frac{2\pi}{L}\left(L_0+\overline L_0-\frac{c}{12}\right)
\en
and
\eq
\hat
H_1=\left(\frac{L}{2\pi}\right)^{1-x_{\varphi}}
\oint\varphi(e^{i\theta})\,d\theta\,.
\en
For  $\lambda$ real in the ADE models we also expect a  $\lambda$--$M$
relation of the form
\eq
M(\lambda)=\kappa |\lambda|^{1/(2-x_{\varphi})}~~;\qquad
|\lambda(M)|=\left(\frac{M}{\kappa}\right)^{2-x_{\varphi}}
\en
with $\kappa$ a model-dependent constant.

Leaving the boundary unperturbed, we have a conformal
boundary condition $\alpha$, with boundary state $\ket{\alpha}$. Set
$g_{\sket{\alpha}}\equiv g^0_{\sket{\alpha}}=\langle \alpha| 0\rangle$ and
$g^{\varphi}_{\sket{\alpha}}=\langle \alpha|
\varphi\rangle$, where $\ket{0}$ and $\ket{\varphi}$ are the states
corresponding to the fields $1$ and $\varphi$. Since the theory is
unitary, $\ket{0}$ is also the unperturbed ground state; and since
$\varphi$ is primary, we have $\bra{0}\varphi\ket{0}=0$.

The aim is to calculate
$\ln\CG_{\sket{\alpha}}(\lambda,L)=\ln\langle \alpha|\Omega\rangle$
where $\ket{\alpha}$ is the unperturbed CFT boundary state, and
$\ket{\Omega}$ is the PCFT vacuum. This will be a power series in
the dimensionless quantity $\lambda L^{2-x_{\varphi}}$; here, we
shall obtain the coefficient of the linear term,
$d_1^{\sket{\alpha}}$. The calculation follows \cite{Dorey:1999cj},
but is a little different (and in fact simpler) because the theory
is unitary, so that the ground state is the conformal vacuum
$\ket{0}$.

First-order perturbation theory implies
\eq
\ket{\Omega}=\ket{0}+\lambda\,{\sum_a}'\Omega_a\ket{\psi_a}+\dots
\en
where the sum is over all states excluding $\ket{0}$\,,
$\langle\psi_a\ket{0}=0$ and
\eq
\Omega_a=
\frac{\bra{\psi_a}\hat H_1\ket{0}}%
{\bra{0}\hat H_0\ket{0}-\bra{\psi_a}\hat H_0\ket{\psi_a}}\,.
\en
Since the theory is unitary,
$\bra{0}\hat H_0\ket{0}-\bra{\psi_a}\hat H_0\ket{\psi_a}=
-\bra{\psi_a}\frac{2\pi}{L}(L_0+\overline L_0)\ket{\psi_a}$.
Using rotational invariance as well,
\eq
\lambda\,{\sum_a}'\Omega_a\ket{\psi_a} =-
\frac{\lambda L^{2-x_{\varphi}}}{(2\pi)^{1-x_{\varphi}}}\,{\sum_a}'\,
\frac{\ket{\psi_a}\bra{\psi_a}\varphi(1)\ket{0}}%
{\bra{\psi_a}L_0{+}\overline L_0\ket{\psi_a}}~.
\en
Hence
\eq
\ket{\Omega}=\ket{0}-
\frac{\lambda L^{2-x_{\varphi}}}%
{(2\pi)^{1-x_{\varphi}}}
\,(1{-}P)
\frac{1}{L_0{+}\overline L_0}(1{-}P)\varphi(1)\ket{0}+\dots
\en
where $P=\ket{0}\bra{0}$ is the projector onto the ground state.
Using the formula
$\frac{1}{L_0{+}\overline L_0}=\int_{0}^{1}q^{L_0{+}\overline
L_0-1}\,dq$\,,
\eq
\langle \alpha \ket{\Omega}=
\langle \alpha \ket{0}-
\frac{\lambda L^{2-x_{\varphi}}}%
{(2\pi)^{1-x_{\varphi}}}
\bra{\alpha}\,(1{-}P)
\int_{0}^{1}\frac{dq}{q}\,
q^{L_0{+}\overline L_0}(1{-}P)\varphi(1)\ket{0}+\dots
\en
Since $\bra{0}\varphi\ket{0}=0$ and
$q^{L_0{+}\overline L_0}\varphi(1)\ket{0}=
q^{L_0{+}\overline L_0}\varphi(1)q^{-L_0{-}\overline L_0}\ket{0}=
q^{x_{\varphi}}\ket{0}$ this last expression simplifies to
\eq
\langle \alpha \ket{\Omega}=
\langle \alpha \ket{0}-
\frac{\lambda L^{2-x_{\varphi}}}%
{(2\pi)^{1-x_{\varphi}}}
\int_{0}^{1}dq\,q^{x_{\varphi-1}}\bra{\alpha }\varphi(q)\ket{0}+\dots
\en
Now $\bra{\alpha}\varphi(q)\ket{0}$ is a disc amplitude, and by
M\"obius invariance it is given by
\eq
\bra{\alpha}\varphi(q)\ket{0}=
g^{\varphi}_{\sket{\alpha}}\,(1-q^{2})^{-x_{\varphi}}\,.
\en
(This is a significant simplification over the nonunitary case
discussed in~\cite{Dorey:1999cj},
where the corresponding amplitude had to be expressed
in terms of hypergeometric functions.)

Taking logarithms,
\bea
\ln \CG_{\sket{\alpha}}(\lambda,L)&=&
\ln g_{\sket{\alpha}}-\frac{\lambda L^{2-x_{\varphi}}}%
{(2\pi)^{1-x_{\varphi}}}\,
\frac{g^{\varphi}_{\sket{\alpha}}}{g_{\sket{\alpha}}}\,
\int_{0}^{1} \; dq\,q^{x_{\varphi}-1}(1-q^{2})^{-x_{\varphi}}+\dots
\nn\\[3pt]
&=&
\ln g_{\sket{\alpha}}+d_1^{\sket{\alpha}} L^{2-x_{\varphi}}+\dots
\eea
and doing the integral,
\eq
d_1^{\sket{\alpha}}=-\frac{1}{2(2\pi)^{1-x_{\varphi}}}\,
\frac{g^{\varphi}_{\sket{\alpha}}}{g_{\sket{\alpha}}}\,
\,{\rm B}(1{-}x_{\varphi},x_{\varphi}/2)
\label{dform}
\en
where ${\rm B}(x,y)=\Gamma(x)\Gamma(y)/\Gamma(x{+}y)$ is the Euler
beta function.

This is a general result. As a non-trivial check we
specialise to the 3-state Potts model,  described by the $A_2$
scattering theory. There are three possible values A, B and C
of the microscopic spin variable, related
by an $S_3$ symmetry. At criticality the model
corresponds to a $c=4/5$ conformal field theory. The primary fields
are the identity  $I$, a doublet of fields $\{\psi,\psi^{\dagger}\}$
of dimensions $\Delta_{\psi}=\overline{\Delta}_{\psi}=2/3$, the energy
operator $\varepsilon$ of dimensions $\Delta_{\varepsilon}=
\overline{\Delta}_{\varepsilon}=2/5$ and a second doublet of fields
$\{\sigma,\sigma^\dagger\}$ with dimensions
$\Delta_{\sigma}=\overline{\Delta}_{\sigma}=1/15$.
The bulk perturbing operator $\varphi$ which leads to the $A_2$
scattering theory is $\varepsilon$, and so
$x_{\varphi}=\Delta_{\varepsilon}+\overline{\Delta}_{\varepsilon}=4/5$.
Boundary conditions and states for the unperturbed model are discussed
in \cite{Cardy:1989ir,Affleck:1998nq}.
One of the three `fixed' boundary states, say $\ket{\mbox{\em A}}$,
can be written in terms of $W_3$-Ishibashi states as
\cite{Cardy:1989ir}
\eq
\ket{\mbox{\em A}}=
\ket{\tilde{I}}\equiv
K\left[\ivec{I}+X \ivec{\varepsilon}+ \ivec{\psi}+
\ivec{\psi^\dagger}+X\ivec{\sigma}+X \ivec{\sigma^\dagger}\right]
\en
where $K^4=(5-\sqrt{5})/30$ and $X^{2}=(1+\sqrt{5})/2$. Hence
\eq
\ln g_{\sket{A}}= \ln K
=-0.5961357674\dots\,,~~~
g^{\varphi}_{\sket{A}}/g_{\sket{A}}=X
= 1.2720196495\dots
\label{gA}
\en
Putting everything
into (\ref{dform}), the CPT prediction for the coefficient of the
first perturbative correction to the $g$-function for fixed boundary
conditions in the three-state Potts model is
\eq
d_1^{\sket{A}}
=-3.011357884\dots
\label{d1}
\en
The 3-state Potts model also admits `mixed' boundary conditions AB, BC
and CA \cite{Saleur:1988zx}. For later use we recall from
\cite{Cardy:1989ir} that the corresponding boundary state is
\eq
\ket{AB}=K\left[ X^{2}
\ivec{I}-X^{-1} \ivec{\varepsilon}+
X^{2} \ivec{\psi}+ X^{2}\ivec{\psi^\dagger}-X^{-1}\ivec{\sigma}-X^{-1}
\ivec{\sigma^\dagger}\right]
\en
so that, for the AB boundary, we find instead
\eq
d_1^{\sket{AB}}= -{1 \over X^4} d_1^{\sket{A}}\,.
\label{d1AB}
\en

The results (\ref{gA}) and  (\ref{d1}) can be compared to the
numerical evaluation of our exact $g$-function result, written in terms
of $\lambda L^{6/5}$ using Fateev's formula~\cite{Fateev:1993av}
\eq
\kappa=\frac{3\Gamma(4/3)}{\Gamma^{2}(2/3)}
(2\pi)^{5/6}(\gamma(2/5)\gamma(4/5))^{5/12}
=4.504307863\dots
\en
where $\gamma(x)=\Gamma(x)/\Gamma(1-x)$. Setting $x=\lambda
L^{6/5}=-(l/\kappa)^{6/5}$ (remembering that $(T-T_c) \propto
\lambda<0$) a fit to our numerical data yielded
\bea
\ln g(l)&=& -0.596135768
 -0.49999991\,l
 -3.0113570\,x \nn\\
&&~ -0.909937\,x^{2} +0.3982\,x^{3}
 +1.0\,x^{4}+\dots
\label{pottsfit}
\eea
which agrees well with~(\ref{gA}) and (\ref{d1}).
The match of the constant term in (\ref{pottsfit})
with $\ln g_{\sket{A}}$ from (\ref{gA})
is guaranteed by our exact formula, and so serves as a check on the
accuracy of our numerics. A calculation of the coefficient of
the irregular (in $x$) term, proportional to
$l$, as in \cite{Dorey:1999cj}, predicts the value $0.5$, again in
good agreement with our numerical results.

\resection{One-parameter families and RG flows}
Just as the minimal reflection factors were tested in section 7, we
do the same now with the one-parameter families of reflection
factors. These reflection factors, $R_a^{\sket{d,C}}$, depend
on the parameter $C$ and are given in (\ref{adeSrick}) and
(\ref{zdef}).

\subsection{ The ultraviolet limit}
\label{UVlim}
If we use the reflection factors as input to calculate the
$g$-function, eq.\,(\ref{exactg}), and take the limit $l \rightarrow 0$ then
\eq
g^{\sket{d}} \equiv g_{\sket{d,C}}(l)|_{l=0} = g_{\bzero} \, T_d~,
\label{GTUV}
\en
where $T_d=T_d(\te)|_{l=0}$ is  a $\te$-independent constant. From
(\ref{YT}) and (\ref{Ye}) we have
\eq
T_d=\sqrt{1+e^{\epsilon_d}}
\en
where  the $\epsilon_d$ values can be found in
\cite{Klassen:1989ui}. For every  ADET theory,
eq.\,(\ref{GTUV}) leads  to a possible CFT
$g$-function value.

\begin{figure} [htbp]
\centering
\includegraphics[width=0.75\textwidth]{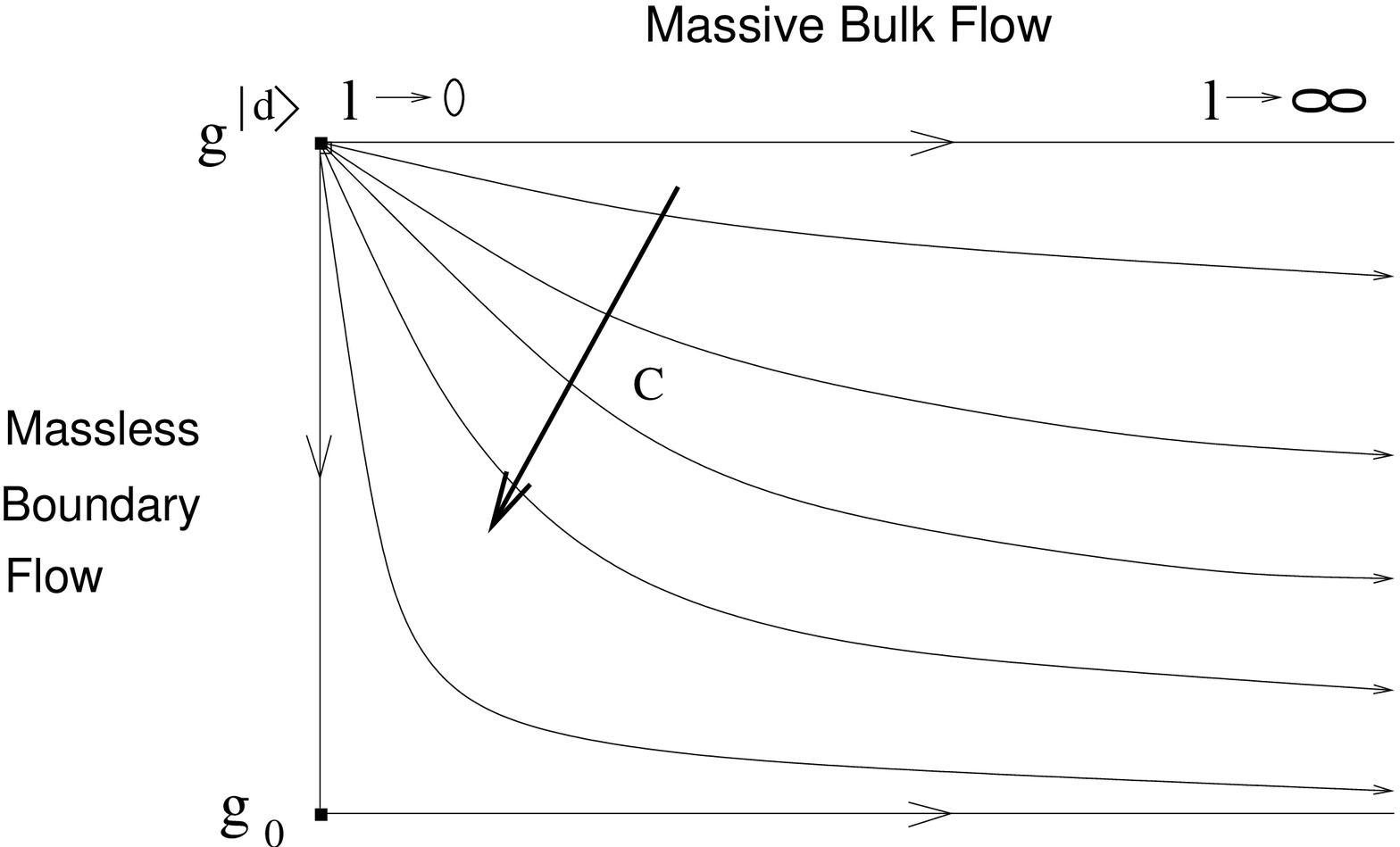}
\caption{\small The expected RG flow pattern}
\label{flow}
\end{figure}

For $M>0$ there will be many massive bulk flows
as the parameter $C$ is varied. However, it should be possible to tune
$C$, as the limit $l\rightarrow 0$ is taken, so as to
give a massless boundary flow between the conformal $g$-function
$g^{\sket{d}}$, corresponding to the UV limit of the
reflection factor
$R_a^{\sket{d,C}}$, and $g$, corresponding to the UV limit of the
minimal reflection factor.
These flows are depicted in figure~\ref{flow}. Note that the UV
$g$-function corresponding to the minimal reflection factors ($g_{\bzero}$ for
the ADE cases and
$g_{(1,1)}$ for $T_r$) is the smallest conformal value in all cases so, by
Affleck and Ludwig's $g$-theorem \cite{Affleck:1991tk}, this is a
stable fixed point of the boundary RG flow.

For the $T_r$ case we find
\eq
g^{\sket{d}} = g_{(1,d+1)} \quad\textrm{for}\quad d=1,\ldots,r.
\label{gbtn}
\en
This is consistent with a simple pattern of flows
\eq
(1,d{+}1)\to (1,1) \quad\textrm{for}\quad d=1,\ldots,r.
\en

The conformal values corresponding to the UV limits of $g^{\sket{d}}$
for the A, D and E cases are given in tables~\ref{tabb4}
and~\ref{tabb5}. Again we expect there to be boundary flows
\eq
g^{\sket{d}}\to g_{\bzero}
\en
in each case. Notice that in many cases the UV $g$-function values
are sums of `simple' CFT values, corresponding to flows from
superpositions of Cardy boundary conditions, driven by
boundary-changing operators.

\begin{table}[htbp]
\begin{center}
\begin{tabular}{|c|c|}
\hline
$A_r$ & $D_r$ \\
\hline
\hline
$g^{\sket{1}} = g^{\sket{r}} = g_{\brho_1}$ &  $g^{\sket{i}} = (i+1)g_{\bzero}$
for $i=1,\ldots,r-2$ \\
$g^{\sket{2}} = g^{\sket{r-1}} = g_{\brho_2}$ &
$g^{\sket{r-1}} = g^{\sket{r}} = g_{\brho_{r-1}}=g_{\brho_r}$  \\
  $\vdots$ &  \\
$g^{\sket{r/2}} = g^{\sket{r/2+1}} = g_{\brho_{r/2}}$ for $r$ even & \\
$g^{\sket{h/2}} = g_{\brho_{h/2}}$ for $r$ odd & \\
\hline
\end{tabular}
\caption{\small UV $g$-function values for A and D models}
\label{tabb4}
\end{center}
\end{table}

\begin{table}[htbp]
\begin{center}
\begin{tabular}{|c|c|c|c|c|}
\hline
 &\multicolumn{2}{|c|}{$E_6$} & $E_7$ & $E_8$ \\
\hline
\hline
$g^{\sket{1}}$ & $g^{\sket{\overline{1}}}$ & $g_{\brho_1}$ & $g_{\brho_1}$ &
$g_{\bzero}+g_{\brho_1}$ \\
\hline
$g^{\sket{2}}$ & & $g_{\bzero}+g_{\brho_2}$ & $g_{\bzero}+g_{\brho_2}$ &
$2g_{\bzero}+g_{\brho_1}$ \\
\hline
$g^{\sket{3}}$ & $g^{\sket{\overline{3}}}$ & $g_{\brho_1}+g_{\brho_2}$ &
$g_{\brho_1}+g_{\brho_3}$ & $2g_{\bzero}+2g_{\brho_1}$ \\
\hline
$g^{\sket{4}}$ & & $g_{\bzero}+g_{\brho_1}+2g_{\brho_2}$ & $g_{\bzero}+2g_{\brho_2}$ &
 $3g_{\bzero}+2g_{\brho_1}$ \\
\hline
$g^{\sket{5}}$ & \multicolumn{2}{|c|}{ } & $g_{\bzero}+3g_{\brho_2}$ &
$5g_{\bzero}+3g_{\brho_1}$ \\
\hline
$g^{\sket{6}}$ & \multicolumn{2}{|c|}{ } & $2g_{\brho_1}+2g_{\brho_3}$ &
$5g_{\bzero}+4g_{\brho_1}$ \\
\hline
$g^{\sket{7}}$ & \multicolumn{2}{|c|}{ } & $3g_{\bzero}+6g_{\brho_2}$ &
$9g_{\bzero}+6g_{\brho_1}$ \\
\hline
$g^{\sket{8}}$ & \multicolumn{2}{|c|}{ } & & $16g_{\bzero}+12g_{\brho_1}$ \\
\hline
\end{tabular}
\caption{\small UV $g$-function values for $E_6$, $E_7$ and $E_8$ models}
\label{tabb5}
\end{center}
\end{table}

The conjectured flow for the three-state Potts model ($A_2$),
$g_{\brho_1} \rightarrow g_{\bzero}$, corresponds to the
`mixed--to--fixed' flow
($AB\rightarrow A$) found by Affleck, Oshikawa and
Saleur~\cite{Affleck:1998nq}, and by
Fredenhagen~\cite{Fredenhagen:2003xf}.

Similarly,  the flow
$g_{\brho_1} \rightarrow g_{\bzero}$ in the tricritical Ising
model ($E_7$) corresponds to  the  `degenerate--to--fixed'
($(d)\rightarrow(-)$ or $(d)\rightarrow(+)$) flows of
\cite{Affleck:2000jv,Fredenhagen:2003xf}. Notice that
$g_{\bzero}+g_{\brho_2} \rightarrow g_{\bzero}$ in $E_7$ matches the CFT
$g$ values for the flow
$(-)\oplus(0+)\rightarrow(-)$ conjectured
in~\cite{Fredenhagen:2003xf}. However this latter flow is driven by
the boundary field with scaling dimension $3/5$, which is
inappropriate for the $E_7$ coset description. A more careful
analysis shows that our flow, which is driven by a boundary field
with scaling dimension $1/10$, must start from either
$(+)\oplus(0+)$ or $(-)\oplus(-0)$, and flow to $(+)$ or $(-)$. This
serves as a useful reminder that the $g$-function values alone do
not pin down a boundary condition, and that this ambiguity can be
physically significant in situations involving superpositions of
boundaries.

\subsection{On the relationship between the UV and IR parameters}

The one-parameter families of boundary
scattering theories introduced above should  describe simultaneous
perturbations of boundary conformal field theories by  relevant
bulk and boundary operators.  The action is
\eq
{\cal A}_{\sket{\alpha}}= {\cal A}^{\bf BCFT}_{\sket{\alpha}}+{\cal
A}^{\bf BULK}+{\cal A}^{\bf BND}_{\sket{\alpha}}
\en
where  ${\cal A}^{\bf BCFT}_{\sket{\alpha}}$ is the
unperturbed boundary CFT action.  We suppose that the boundary
condition is imposed
at $x=0$\,; in general it might correspond to a  superposition of
$n_{\sket{\alpha}}$ Cardy states.
Denoting these by $\ket{c}$, $c=1,2,\dots,n_{\sket{\alpha}}$,
the boundary is in the state
\eq
\ket{\alpha} =\sum_{c=1}^{n_{\sket{\alpha}}} n_{\sket{c}} \ket{c}\,,
\en
with $n_{\sket{c}} \in \NN$. The bulk perturbing term is
\eq
{\cal A}^{\bf BULK}= \lambda \int_{-\infty}^{0} dx
\int_{-\infty}^{\infty} dy \; \varphi(x,y)
\en
while the boundary perturbing  part is
\eq
{\cal A}^{\bf BND}_{\sket{\alpha}}= \sum_{c,d=1}^{n_{\sket{\alpha}}}
\mu_{\sbAB{c}{d}} \int_{-\infty}^{\infty} dy \;
\phi_{\sbAB{c}{d}}(y)~.
\en
The operators
$\phi_{\sbAB{c}{c}}$ live on a single Cardy
boundary, while the $\phi_{\sbAB{c}{d}}$ with
$c \ne d$  are boundary changing operators.
Since
$\phi_{\sbAB{c}{d}}=\phi_{\sbAB{d}{c}}^{\dagger}$, in  a unitary theory we
also  expect (see for example~\cite{Graham:2001pp})
\eq
 \mu_{\sbAB{c}{d}}=\mu^{*}_{\sbAB{d}{c}}~.
\en

Using periodicity arguments to analyse
the behaviour of the ground-state energy on a strip with perturbed
boundaries as in~\cite{Zamolodchikov:1991et}
and \cite{Dorey:1997yg}
(see also (\ref{exep}) and
(\ref{tcoeff}) below), one can  argue that the scaling dimensions
of the fields $\phi_{\sbAB{c}{d}}$ must, in these integrable cases,
be half that of the bulk
perturbing operator $\varphi$ :
\eq
x_{\phi}={x_{\varphi} \over 2}= {2 \over h+2}
\label{dim}
\en
 for the ADE systems
and
\eq
x_{\phi}={x_{\varphi} \over 2}= {2-h \over h+2}
\en
 for the
$T_r$ models. The  results recorded in eq.\,(\ref{gbtn}) and
tables~\ref{tabb4} and~\ref{tabb5} provide information about the
conformal boundary conditions associated with the one-parameter
families of reflection factors.
We see that for $T_r$ and $A_r$
the boundary is always in a pure Cardy state, while for
$D_r$, $E_6$ and $E_7$ this is true only in one case per model,
and in the $E_8$-related theories the
UV boundary always corresponds to a non-trivial
superposition of the states $\ket{\pm}$ and $\ket{\bf free}$. This
observation fits nicely with the conformal field theory
results for the Ising model~\cite{Cardy:1989ir}: from (\ref{dim}) we
see that for the $E_8$ coset description the integrable
boundary perturbation must have dimension
$x_{\phi}=1/16$, and indeed  the only boundary operators
with this dimension in the Ising model are
$\phi_{\sbAB{\pm}{\bf free}}$ and   $\phi_{\sbAB{\bf
free}{\pm}}$.

For simplicity, we shall only discuss the cases involving a single
Cardy boundary, where
\eq
{\cal A}^{\bf BND}_{\sket{\alpha}}=  \mu
\int_{-\infty}^{\infty} dy
\;
\phi(y)~.
\en
As in the  Ising and Lee-Yang examples of \cite{Ghoshal:1993tm,
Dorey:1997yg}, a simple formula is expected to link  the couplings
$\lambda$ and $\mu$  of  bulk and boundary fields
to the parameter $C$ in the reflection factors. However, without a precise
identification of the  operator $\phi$
it is  hard to see how such relation can be determined. Even so, a
general argument combined with a numerically-supported
conjecture allows the relation
formula to be fixed up to
a single overall dimensionless  constant. This goes as
follows. From section~\ref{sprope} we know  that in all cases
\eq
{\cal G}_{\sket{d,C}}^{(0)}(l)={\cal G}^{(0)}(l)\,T_{d}(i
\fract{\pi}{h} C)
\label{GT}
\en
where $T_d$ is the TBA-related T-function, and ${\cal G}^{(0)}(l)$ is
the CPT ${\cal G}$-function corresponding to the minimal reflection factor,
for which there is no boundary perturbation.
In addition, $T_d(\te)
\equiv T_d(\te,l)$ is even in $\theta$, $T_d(\te)=T_d(-\te)$, and
periodic,
 $T_d(\te+i \pi \fract{h+2}{h})=T_d(\te)$, and so it
can  be Fourier expanded as
\eq
T_d(i\fract{\pi}{h} C) = c_0(l)+ \sum_{k=1}^{\infty} c_k(l)
\cos\left({2\pi k
\over h+2} C  \right)\,.
\label{exep}
\en
We can now  use the observation of \cite{Dorey:1997rb} that
$T_a(\te,l)$ admits an expansion with finite domain of convergence
in the pair of variables
$a_{\pm}=(l e^{\pm \te})^{\fract{2h}{h+2}}$
to see that
\eq
c_0(l)=c_0 +  O(l^{\fract{4h}{h+2}} )~,~c_1(l)=c_1
\, l^{\fract{2h}{h+2}} +  O(l^{\fract{4h}{h+2}})~.
\label{tcoeff}
\en
We also know that the minimal ${\cal G}$-function has an expansion
\eq
\ln {\cal G}^{(0)}(l)= \ln {\cal G}^{(0)} + \sum_{k=1}^{\infty} g_k
l^{k(2-2x_{\phi})}
\label{cptres}
\en
while the conformal perturbation theory expansion of ${\cal
G}_{\sket{d,C}}^{(0)}$ has the form (see~\cite{Dorey:1999cj})
\eq
\ln {\cal G}_{\sket{d,C}}^{(0)}(\lambda,\mu,L)=
\sum_{m,n=1}^{\infty} c_{mn}
(\mu L^{1-x_{\phi}})^m
(\lambda L^{2-2x_{\phi}})^n.
\label{cptresg}
\en
Comparing (\ref{GT}) -- (\ref{cptres}) with (\ref{cptresg})
we conclude that, so long as $c_{10}\neq 0$,
the  relationship between
$C$  and $\mu$ must have the form
\eq
\mu ={\hat \mu}_0 \cos\left({2\pi \over h+2} C  \right)
M^{\fract{2h}{h+2}}={\hat \mu}_0 \cos\left({2\pi \over h+2} C
\right) M^{1-x_{\phi}}
\label{hm}
\en
where ${\hat \mu}_0$ is an unknown dimensionless constant. However
the result  (\ref{hm}) can only be correct if $1-x_{\phi}=2h/(h+2)$.
This is true only in the non-unitary $T_r$ models, and indeed
it  reproduces the Lee-Yang result of
\cite{Dorey:2004xk} when  specialised to
$T_1$. For the ADE theories, $1-x_{\phi}=h/(h+2)$  and
we conclude that $c_{10}$, the first
$\mu$-dependent
correction to ${\cal G}$, must be zero. (This is not surprising since
in a unitary CFT this  correction is
proportional to $\vev{\phi}_{\sket{\alpha}}^{\bf disk}=0$.) The
first contribution  is then at order
$O(\mu^{2})=O(M^{2-2x_{\phi}})$, and at this order there is
an overlap between the expansions of
$T_d(\te,l)$ and
${\cal G}^{(0)}(l)$. This leads to the less-restricted result
\eq
\mu^{2} ={\hat k}_0 \left( {\hat z}+\cos\left({2\pi \over h+2} C
\right) \right) M^{\fract{2h}{h+2}}=
{\hat k}_0 \left( {\hat z}-\cos\left({2\pi \over h+2} C
\right) \right)   M^{2-2 x_{\phi}}
\label{hm1}
\en
where now both ${\hat k}_0$ and ${\hat z}$ are unknown constants. If
we now consider the  Ising model, then the $\mu-C$ formula is known
\cite{Ghoshal:1993tm}. Written in terms of $C$ it becomes
\eq
\left(h^{\hbox{ref}\cite{Ghoshal:1993tm}} \right)^{2}=
\mu^{2}=2M \left(1+\cos( \fract{\pi}{2} C)
\right)
\Rightarrow
\mu =2 \sqrt{M}
\cos(\fract{\pi}{4} C)\,.
\en
Thus the boundary magnetic field is an even function of
$C$. It  is then  tempting to conjecture  that ${\hat z}=1$ for all
the $g_{\sket{1,C}}$ cases in the  $A_r$ models, and, to preserve the
perfect square property, that ${\hat z}$ is
either $1$ or $-1$ in all other ADE single boundary condition
situations:
\bea
\!\!\!\!\!\!\!\!\!
 {\hat z}=~1&:&\mu ={\hat \mu}_0
\cos\left(\frac{\pi\,C}{h{+}2}\right) M^{1-x_{\phi}}
\,\leftrightarrow\,
 \frac{\mu}{\sqrt{\lambda}} ={\hat \mu}_0\kappa^{1-x_{\phi}}
\cos\left(\frac{\pi\,C}{h{+}2}\right) ;~~~
\label{hmade} \\[3pt]
\!\!\!\!\!\!\!\!\!
 {\hat z}=-1&:&\mu = {\hat \mu}_0
\sin\left(\frac{\pi\,C}{h{+}2}\right) M^{1-x_{\phi}}
\,\leftrightarrow\,
 \frac{\mu}{\sqrt{\lambda}} ={\hat \mu}_0\kappa^{1-x_{\phi}}
\sin\left(\frac{\pi\,C}{h{+}2}\right) .~~~
\label{hmade1}
\eea
We now return to the physical picture of flows parametrised by $C$
depicted in figure~\ref{flow}. At $\lambda=0$ the bulk mass is zero,
and the only scale in the problem is that induced by the boundary
coupling
$\mu$. The massless boundary flow down the left-hand edge of the
diagram therefore corresponds to varying $|\mu|$ from $0$ to
$\infty$. If $\lambda$ is instead kept finite and nonzero while
$|\mu|$ is sent to infinity, the flow will collapse onto the lower
edge of the diagram, flowing from $g_0$ in the UV. For the
$g$-function calculations to reproduce this behaviour, the
reflection factor should therefore reduce to its minimal version as
$|\mu|\to\infty$.
For (\ref{hmade}), $|\mu|\to\infty$ corresponds to $C\to i\infty$,
which does indeed reduce the reflection factor as required.
On the other hand, taking
$|\mu|\to\infty$ in (\ref{hmade1}),
requires $C\to (h{+}2)/2 + i\infty$. While the reduction is again
achieved in the limit, real analyticity of the reflection factors is
lost at intermediate values of $\mu$. For this reason option
(\ref{hmade}) might be favoured, but more detailed work will be
needed to make this a definitive conclusion.

In fact, the proposal (\ref{hmade}) can be checked at $\mu=0$ in the
3-state Potts model ($h=3$), as follows. Consider the results~(\ref{d1})
and (\ref{d1AB}) and set
\eq
\delta_1=- \kappa^{-6/5}\left(d_1^{\sket{AB}}-d_1^{\sket{A}}\right )=
\kappa^{-6/5}\left(1+{1 \over X^4}
\right)d_1^{\sket{A}} = -0.683763720\dots
\en
According to the conclusions  of section~\ref{UVlim} and
eq.\,(\ref{hmade}),
\eq
{\cal G}_{\sket{AB}}^{(0)}(l)={\cal G}_{\sket{A}}^{(0)}(l)
T_{1}(\fract{\pi}{3} C)|_{C=5/2}~~,
\en
and $\delta_1$ should match the coefficients
$t_1$ of $l^{6/5}$ in the expansion of the function
$T_1(\te,l)|_{\te=0}$ about  $l=0$. Noticing
that $T_1(\te,l)=T_2(\te,l)=T_{\bf LY}(\te,l)$ we can use
table 6 of \cite{Dorey:1999cj}: $\ln T_{\bf LY}(i \pi(b+3)/6)= \ep(i
\pi(b+3)/6)$,
$C=5/2$ corresponds to
$b=2$,
$h M^{-6/5}=h_c=-0.6852899839$, and we find
\eq
t_1 \sim 0.9977728224 \, h_c=-0.683763721\dots
\en
which within  numerical accuracy is equal to $\delta_1$.

The study of the cases where the theory is perturbed by  boundary
changing operators is more complicated.  Our results are very
preliminary,  and we have decided to postpone the discussion to  another
occasion.

\resection{Conclusions}
Since the initial studies of integrable boundary scattering theories,
surprisingly little progress has been made on the
identification of the many known boundary reflection factors with
particular perturbations of conformal field theories. One aim
of this paper has been to use the recently-discovered
exact expression for the ground-state degeneracy  $g$ to begin to
fill this
gap. Working within the framework of the minimal ADET models we were
able to identify  a special class of reflection factors possessing a
very simple conformal field theory interpretation: in the
ultraviolet limit they describe the reflection of massless particles
against a wall with fixed  boundary conditions
$\ket{i}$ matching the low-temperature vacua.

A collection of
one-parameter families of amplitudes was also proposed, and partially
justified by a suggested quantum group
reduction of the boundary sine-Gordon model at
$\beta^{2}=16 \pi/(2r+3)$.

The consistency  of these novel  families of boundary scattering
theories was supported by our $g$-function  calculations: they
appear to correspond to  perturbations of Cardy boundary states, or
of superpositions of such states, by relevant boundary operators,
with each boundary flow ending at a fixed boundary condition
$\ket{i}$.

While our results indicate that these  models all have
interesting interpretations as  perturbed boundary conformal field
theories, the  checks performed so far should be considered as preliminary,
and more work will be needed to put these results on a firmer footing.
Two immediate open problems are
fixing completely  the boundary UV/IR relations, and determining the
boundary bound-state content of each ADET  model.
The effort involved should be rewarded both by
the well-established role in condensed matter physics taken by some of
these theories, and by the mathematical insights into general
integrable boundary theories their further study should bring.

\subsection*{Acknowledgements}

\noindent
We would like to thank  Daniel Roggenkamp for his helpful
explanations of coset conformal field theories, and Zoltan Bajnok,
Davide Fioravanti, Gerard Watts, Cristina Zambon and Aliosha
Zamolodchikov for useful discussions. PED and ARL thank APCTP,
Pohang, for hospitality during the Focus Program `Finite-Size
Technology in Low Dimensional Quantum Field Theory', where much of
this material was presented, and INFN, Torino University for
hospitality in the final stage of the project. The work was partly
supported by the EC network ``EUCLID", contract number
HPRN-CT-2002-00325, and partly by a NATO grant PST.CLG.980424. PED
was also supported in part by JSPS/Royal Society grant and by the
Leverhulme Trust, and the work by CR was partly supported by the
Korea Research Foundation 2002-070-C00025, and by the Science
Research Center Program of the Korea Science and Engineering
Foundation through the Center for Quantum Spacetime (CQUeST) of
Sogang University with grant number R11-2005-021.
\appendix
\resection{Diagonal $\hat{\mathfrak{g}}_1\times
\hat{\mathfrak{g}}_1/\hat{\mathfrak{g}}_2$ coset data}
\label{coset}
In general, to extract the $\hat{\mathfrak{g}}/\hat{\mathfrak{h}}$
coset conformal theory from the $\hat{\mathfrak{g}}$ WZW model, we need
to decompose the representations, $\bmu$, of
$\hat{\mathfrak{g}}$ into a direct sum of representations, $\bnu$, of
$\hat{\mathfrak{h}}$:
\eq
\bmu \mapsto \bigoplus_{\bnu}b_{\bmu\bnu}\bnu.
\en
This decomposition corresponds to the character identity
\eq
\chi_{\bmu}(\tau)=\tilde{\chi}_{\bnu}(\tau)b_{\bmu \bnu}(\tau)
\en
where $\chi_{\bmu}$ and $\tilde{\chi}_{\bnu}$ are the characters of the
$\hat{\mathfrak{g}}$
and $\hat{\mathfrak{h}}$ representations $\bmu$ and $\bnu$ respectively and
the branching function, $b_{\bmu\bnu}(\tau)$, is the character
of the coset theory.
Let $\Pi$ denote the projection matrix giving the explicit projection
of every weight of $\mathfrak{g}$ onto a weight of
$\mathfrak{h}$. Clearly for a coset character to be non zero we must have
\eq\label{rule1}
\Pi \bmu_0 -\bnu_0 \in \Pi Q
\en
where $Q$ is the root lattice of $\mathfrak{g}$. For our diagonal cosets,
since
$\Pi(Q\oplus Q)=Q$ this selection rule is particularly simple:
\eq
\bmu_0 +\bnu_0 -\brho_0 \in Q
\label{rule11}
\en
where $\bmu_0,\bnu_0$ and $\brho_0$ are weights of $\mathfrak{g}$.

The group of outer automorphisms of $\hat{\mathfrak{g}}$,
$O(\hat{\mathfrak{g}})$, permutes
the fundamental weights in such a way as to leave the extended
Dynkin diagram invariant. The action of an element, $A\in
O(\hat{\mathfrak{g}})$, on a modular S-matrix element of $\hat{\mathfrak{g}}$
at some level $l$ is
\eq \label{Saut}
S_{(A\bmu)\bmu'}=S_{\bmu\bmu'}e^{2\pi i(A\bLambda_0,\bmu_0')}.
\en
The modular S-matrix for a diagonal coset theory has the form
\eq
S_{\{\bmu,\bnu;\brho\}\{\bmu',\bnu';\brho'\}}=S_{\bmu\bmu'}^{(1)}
S_{\bnu\bnu'}^{(1)}S_{\brho\brho'}^{(2)*}.
\en
Under the $O(\hat{\mathfrak{g}})$ action it transforms as
\bea
S_{\{A\bmu,A\bnu;A\brho\}\{\bmu',\bnu';\brho'\}} &=& e^{2\pi
i(A\bLambda_0,\bmu_0'+\bnu_0'-\brho_0')}
S_{\{\bmu,\bnu;\brho\}\{\bmu',\bnu';\brho'\}} \\
&=& S_{\{\bmu,\bnu;\brho\}\{\bmu',\bnu';\brho'\}} \nn
\eea
since $\bmu_0'+\bnu_0'-\brho_0'\in Q$ \cite{Gepner:1989jq}. This suggests that
$S$ is a degenerate matrix, which cannot be the case since it
represents a modular transformation. We must therefore remove this
degeneracy by identifying the fields
\eq \label{id}
\{A\bmu,A\bnu;A\brho\}\equiv\{\bmu,\bnu;\brho\}\quad \forall A\in
O(\hat{\mathfrak{g}}).
\en
For these diagonal cosets, the orbit of every element $A$ of
$O(\hat{\mathfrak{g}})$ has the same order, $N$, which is simply the order of
the global symmetry group of the model. With this field
identification, we find that every field in the theory appears with
multiplicity $N$. To remedy this, the partition function must be
divided by this multiplicity, which has the effect of introducing
$N$ into the coset modular S-matrix
\cite{Schellekens:1989uf}:
\eq
S_{\{\bmu,\bnu;\brho\}\{\bmu',\bnu';\brho'\}}=NS_{\bmu\bmu'}^{(1)}
S_{\bnu\bnu'}^{(1)}S_{\brho\brho'}^{(2)}.
\en
More information on field identifications can be found, for example, in
\cite{Ahn:1989re}.

For the level 1 representations of simply laced affine Lie algebras, the
characters have a particularly simple
form\,\cite{Frenkel:1980rn,Kac:1984mq}\,:
\eq
\chi_{\bmu}(q)=\frac{1}{\eta^{\rg}(q)}\theta_{\bmu}(q)
\en
where $q=\exp(2\pi i\tau)$ and the generalised $\theta$ and
Dedekind $\eta$ functions are
\bea
\theta_{\bmu}(q) &=& \sum_{\bmu_0\in Q}q^{\frac{1}{2}(\bmu_0,\bmu_0)} \\
\eta(q) &=& q^{\frac{1}{24}}\prod_{n=1}^\infty (1-q^n).
\eea
Under the modular transformation $\tau\rightarrow -\frac{1}{\tau}$ the theta
function transforms as
\eq
\theta_{\bmu}\rightarrow (-i\tau)^{\rg/2}\frac{1}{\sqrt{|P/Q|}}
\sum_{\bnu_0\in P/Q} e^{-2\pi i(\bnu_0,\;\bmu_0)}\theta_{\bnu}
\en
where $P$ and $Q$ are the weight lattice and root lattice respectively
\cite{Kac:1984mq,Kac:1990gs}, and the $\eta$ function becomes
\eq
\eta(-\frac{1}{\tau})=(-i\tau)^{\frac{1}{2}}\eta(\tau).
\en
The modular S-matrix for level 1 is
therefore
\eq
S^{(1)}_{\bmu \bnu} = \frac{1}{\sqrt{|P/Q|}}\,e^{-2\pi i(\bnu_0,\bmu_0)}.
\en
For the coset description of the $g$-function, eq.(\ref{gcoset}), we
need to compute
$S^{(1)}_{\bzero\bmu}$ which is simply
\eq \label{Slevel1}
S^{(1)}_{\bzero\bmu} = \frac{1}{\sqrt{|P/Q|}}.
\en
It is important to note that $P/Q$ is isomorphic to the centre of
the group under consideration
\cite{Bernard:1986xy,Ahn:1989re}. The group of field
identifications of the coset is also isomorphic to the centre of
this group so consequently eq.(\ref{Slevel1}) becomes
\eq
S^{(1)}_{\bzero\bmu}=\frac{1}{\sqrt{N}}\quad,\quad\textrm{for all $\bmu$}.
\en
The $g$-function can now be written in terms of the level 2 modular S
matrices only:
\eq \label{Ag}
g_{\{\bmu,\bnu;\brho\}} =
\frac{S^{(2)}_{\bzero\brho}}{\sqrt{S^{(2)}_{\bzero\bzero}}}.
\en

As we can see from (\ref{Ag}), identifying the conformal
$g$-function value will only pin down the level 2 representation.
For $E_8$, since there is only one possible level 1 representation
(with Dynkin labels $n_i=0$, $i=1,\ldots,8$), by specifying the
level 2 representation we fix the coset field. However, for other
cases, although the coset selection and identification rules
(\ref{rule11}),(\ref{id}) do constrain the possible coset
representations we are still left with some ambiguity in general.
For example, for the $A_r$ cases the selection and identification
rules are quite simple and the possible cosets, given a fixed level
2 weight, are shown in table~\ref{Ar}. The notation is as follows:
$\bmu_j$, $\bnu_j$ are level 1 weights and $\brho_j$ is the level 2 weight
with Dynkin labels $n_i=1$ for $i=j$ and $n_i=0$, $i=1,\ldots,r$ otherwise.
The labels $\bmu_0$, $\bnu_0$ and $\brho_0$ now represent $\bzero$ with Dynkin
labels $n_i=0$, $i=1,\ldots,r$.

\begin{table}[htbp]
\small
\begin{center}
\begin{tabular}{|c|c|c|c|}
\hline
&&&\\[-9pt]
Fixed level 2 weight & Level 1 weight  & Level 1 weight &
Number of coset\\
$\brho$ & $\bmu$ &  $\bnu$ &fields  $\{\bmu,\bnu;\brho\}$ \\[3pt]
\hline
\hline
&&&\\[-9pt]
$\bzero$ & $\bzero$ & $\bzero$ &  \\
 & $\bmu_i$ & $\bnu_{r+1-i}$, $i=1,\ldots r$ & $r+1$ \\[3pt]
\hline
&&&\\[-9pt]
$\brho_j$ & $\bmu_i$ & $\bnu_{j-i}$, $i=0,\ldots j$ & \\
$j=1,\ldots [(r+1)/2]$ & $\bmu_{j+k}$ & $\bnu_{r+1-k}$, $k=1,\ldots
r{-}j$ &
$r+1$ \\[3pt]
\hline
\end{tabular}
\caption{\small $A_r$ coset fields, indicating the number of distinct
fields for each level 2 weight; in the first column,
$[x]$ denotes the integer part of $x$}
\label{Ar}
\end{center}
\end{table}

The coset fields for the $A_2$ (three-state Potts model) and $E_7$ (tricritical
Ising model) cases, along with the corresponding
boundary condition labels from \cite{Affleck:1998nq} and \cite{Chim:1995kf}
respectively, are given in tables~\ref{A2} and~\ref{E7} as concrete
examples.

\begin{table}[ht]
\small
\begin{center}
\begin{tabular}{|c|c|c|}
\hline
\multicolumn{3}{|c|}{} \\[-9pt]
\multicolumn{3}{|c|}{$A_2$}  \\[3pt]
\hline
\hline
&&\\[-9pt]
Coset field & Boundary label & Level 2 weight label  \\
 &from \cite{Affleck:1998nq} & from table~\ref{tabb2} \\[3pt]
\hline
&&\\[-9pt]
$\{[1,0,0],[1,0,0];[2,0,0]\}$ & $A$ &  \\
$\{[0,0,1],[0,1,0];[2,0,0]\}$ & $B$ & $\bzero$ \\
$\{[0,1,0],[0,0,1];[2,0,0]\}$ & $C$ & \\[3pt]
\hline
&&\\[-9pt]
$\{[1,0,0],[0,1,0];[1,1,0]\}$ & $AB$ & \\
$\{[0,0,1],[0,0,1];[1,1,0]\}$ & $BC$ & $\brho_1$ \\
$\{[0,1,0],[1,0,0];[1,1,0]\}$ & $AC$ & \\[3pt]
\hline
\end{tabular}
\caption{\small $A_2$ coset fields with the corresponding
boundary labels and level 2 weight labels from table~\ref{tabb2}.
The weights are given in terms of Dynkin labels $[n_0,n_1,n_2,\ldots]$}
\label{A2}
\end{center}
\end{table}

\begin{table}[tbp]
\small
\begin{center}
\begin{tabular}{|c|c|c|}
\hline
\multicolumn{3}{|c|}{} \\[-9pt]
\multicolumn{3}{|c|}{$E_7$} \\[3pt]
\hline
\hline
&&\\[-9pt]
Coset field & Boundary label & Level 2 weight label \\
 & from \cite{Chim:1995kf} & from table~\ref{tabb2} \\[3pt]
\hline
&&\\[-9pt]
$\{[1,0,\ldots,0],[1,0,\ldots,0],[2,0,\ldots,0]\}$ & $(-)$ & $\bzero$ \\
$\{[0,1,0,\ldots],[0,1,0,\ldots];[2,0,\ldots,0]\}$ & $(+)$ & \\[3pt]
\hline
&&\\[-9pt]
$\{[1,0,\ldots,0],[0,1,0,\ldots];[1,1,0,\ldots,0]\}$ & $(d)$ &
$\brho_1$ \\[3pt]
\hline
&&\\[-9pt]
$\{[1,0,\ldots,0],[1,0,\ldots,0];[0,0,1,0,\ldots]\}$ & $(-0)$ & $\brho_2$ \\
$\{[0,1,0,\ldots],[0,1,0,\ldots];[0,0,1,0,\ldots]\}$ & $(0+)$ &
\\[3pt]
\hline
&&\\[-9pt]
$\{[1,0,\ldots,0],[0,1,0,\ldots];[0,0,0,1,0,\ldots]\}$ & $(0)$ &
$\brho_3$ \\[3pt]
\hline
\end{tabular}
\caption{\small $E_7$ coset fields with the corresponding
boundary labels and level 2 weight labels from
table~\ref{tabb2}. The weights are given in terms of Dynkin labels $[n_0,n_1,n_2,\ldots]$}
\label{E7}
\end{center}
\end{table}

It is useful to note that for the $A_r$, $D_r$, $E_6$ and $E_7$ models,
$S_{\bzero\bzero}=S_{\bzero\brho}$ only when $\brho=A\bzero$ for some
$A\in O(\hat{\mathfrak{g}})$~\cite{Fuchs:1990wb} and for each such $\brho$
there is a unique coset field, so the number of fields with $g$-function equal
to $g_{\{\bmu,\bnu,\bzero\}}$ is equal to the size of the orbit of $\bzero$.
On the other hand, $E_8$ has no diagram symmetry, but it is also exceptional
in that $S_{\bzero\bzero}=S_{\bzero\brho_2}$
where $\brho_2$ has Dynkin labels
$n_2=1$, all other $n_i=0$. Physically,
this is to be expected as the two fields correspond to the two
fixed boundary conditions $(-)$ and $(+)$, which clearly
must have equal $g$-function values.  (This exceptional equality is
discussed from a more mathematical perspective in, for
example,~\cite{Fuchs:1990wb}.)

{}From (\ref{Saut}) it is clear that
$S_{\bzero(A\brho)}=S_{\bzero\brho}$ is also true for
$\brho\neq\bzero$. For the $A$ and $E$ models at level 2, we find
that these are the only cases where
$S_{\bzero\brho_i}=S_{\bzero\brho_j}$; for the $D_r$ models there is
more degeneracy.


\newpage

\begin{thebibliography}{99}
\raggedright
\parskip 1pt

%
%
\bibitem{bologna}
  Al.B.~Zamolodchikov, talk at the 4th Bologna workshop on
  conformal and integrable models, July 3, 1999 [unpublished].
%
\bibitem{Dorey:1997yg}
P.~Dorey, A.~Pocklington, R.~Tateo and G.M.T.~Watts,
{\em `TBA and TCSA with boundaries and excited states',}
Nucl.\ Phys.\ B {\bf 525} (1998) 641
[arXiv:hep-th/9712197].
%
\bibitem{Bajnok:2001ug}
Z.~Bajnok, L.~Palla and G.~Takacs,
{\em `Spectrum and boundary energy in boundary sine-Gordon theory',}
Nucl.\ Phys.\ B {\bf 622} (2002) 565
[arXiv:hep-th/0108157].
%
\bibitem{Corrigan:2000fm}
E.~Corrigan and A.~Taormina, {\em `Reflection factors and a two-parameter
family of boundary bound states  in  the sinh-Gordon model',}
J.\ Phys.\ A {\bf 33}, 8739 (2000)
  [arXiv:hep-th/0008237].
%
\bibitem{Kormos:2002ya}
  M.~Kormos and L.~Palla,
{\em `Some semi-classical issues in boundary sine-Gordon model',}
  J.\ Phys.\ A {\bf 35}, 5471 (2002)
  [arXiv:hep-th/0201230].
\bibitem{Zamolodchikov:1989cf}
Al.B.~Zamolodchikov, {\em `Thermodynamic Bethe ansatz in
relativistic models. Scaling three state Potts and Lee-Yang
models',} Nucl.\ Phys.\ B {\bf 342}, 695 (1990).
%
%
\bibitem{Klassen:1989ui}
T.R.~Klassen and E.~Melzer, {\em `Purely elastic scattering theories
and their ultraviolet limits',} Nucl.\ Phys.\ B {\bf 338} (1990)
485.
%
%
\bibitem{Affleck:1991tk}
I.~Affleck and A.W.W.~Ludwig, {\em `Universal noninteger `ground
state degeneracy' in critical quantum systems',} Phys.\ Rev.\ Lett.\
{\bf 67} (1991) 161.
%
\bibitem{LeClair:1995uf}
A.~LeClair, G.~Mussardo, H.~Saleur and S.~Skorik,
{\em `Boundary energy and boundary states in integrable quantum field
theories',}
Nucl.\ Phys.\ B {\bf 453} (1995) 581
[arXiv:hep-th/9503227].
%
\bibitem{Dorey:1999cj}
P.~Dorey, I.~Runkel, R.~Tateo and G.M.T.~Watts,
{\em `g-function flow in perturbed boundary conformal field
theories',}
Nucl.\ Phys.\ B {\bf 578} (2000) 85
[arXiv:hep-th/9909216].
%
\bibitem{Dorey:2004xk}
P.~Dorey, D.~Fioravanti, C.~Rim and R.~Tateo, {\em `Integrable
quantum field theory with boundaries: the exact g-function',} Nucl.\
Phys.\ B {\bf 696} (2004) 445 [arXiv:hep-th/0404014].
%
\bibitem{Braden:1989bu}
H.W.~Braden, E.~Corrigan, P.~Dorey and R.~Sasaki, {\em `Affine Toda
field theory and exact S matrices',} Nucl.\ Phys.\ B {\bf 338}
(1990) 689.
%
\bibitem{Fateev:1990hy}
V.A.~Fateev and A.B.~Zamolodchikov, {\em `Conformal field theory and
purely elastic S matrices',} Int.\ J.\ Mod.\ Phys.\ A {\bf 5}, 1025
(1990).
%
%
\bibitem{Dorey:1990xa}
P.~Dorey, {\em `Root systems and purely elastic S matrices',} Nucl.\
Phys.\ B {\bf 358} (1991) 654.
%
\bibitem{Dorey:1991zp}
P.~Dorey,
{\em `Root systems and purely elastic S matrices. 2',}
Nucl.\ Phys.\ B {\bf 374} (1992) 741
[arXiv:hep-th/9110058].
%
\bibitem{Fring:1991gh}
A.~Fring and D.I.~Olive, {\em `The fusing rule and the scattering
matrix of affine Toda theory',} Nucl.\ Phys.\ B {\bf 379} (1992)
429.
%
\bibitem{Ravanini:1992fi}
F.~Ravanini, R.~Tateo and A.~Valleriani,
{\em `Dynkin TBAs'},
Int.\ J.\ Mod.\ Phys.\ A {\bf 8} (1993) 1707
[arXiv:hep-th/9207040].
%
\bibitem{Zamolodchikov:1989fp}
A.B.~Zamolodchikov, {\em `Integrals of motion and S matrix of the
(scaled) T = T(C) Ising model with magnetic field',} Int.\ J.\ Mod.\
Phys.\ A {\bf 4}, 4235 (1989).
%
\bibitem{Zamolodchikov:1990jh}
A.B.~Zamolodchikov, {\em `Integrable field theory from conformal
field theory',} Proceedings of Taniguchi Symposium, Kyoto (1988).
%
%
\bibitem{Dorey:1996gd}
P.~Dorey, {\em `Exact S matrices',} in the proceedings of the 1996
E\"otv\"os Graduate School [arXiv:hep-th/9810026].
%
%
\bibitem{Smirnov:1989hh}
F.A.~Smirnov, {\em `The perturbated $c<1$ conformal field theories
as reductions of sine-Gordon model',} Int.\ J.\ Mod.\ Phys.\ A {\bf
4} (1989) 4213.
%
\bibitem{Cardy:1989fw}
J.L.~Cardy and G.~Mussardo, {\em `S matrix of the Yang-Lee edge
singularity in two-dimensions',} Phys.\ Lett.\ B {\bf 225} (1989)
275.
%
\bibitem{Freund:1989jq}
P.G.O.~Freund, T.R.~Klassen and E.~Melzer, {\em `S matrices for
perturbations of certain conformal field theories',} Phys.\ Lett.\ B
{\bf 229}, 243 (1989).
%
\bibitem{Dorey:1997rb}
P.~Dorey and R.~Tateo, {\em `Excited states in some simple perturbed
conformal field theories',} Nucl.\ Phys.\ B {\bf 515} (1998) 575
[arXiv:hep-th/9706140].
%
%
\bibitem{Fring:1993mp}
A.~Fring and R.~Koberle, {\em `Factorized scattering in the presence
of reflecting boundaries',} Nucl.\ Phys.\ B {\bf 421} (1994) 159
[arXiv:hep-th/9304141].
%
\bibitem{Ghoshal:1993tm}
S.~Ghoshal and A.B.~Zamolodchikov, {\em `Boundary S matrix and
boundary state in two-dimensional integrable quantum field theory',}
Int.\ J.\ Mod.\ Phys.\ A {\bf 9} (1994) 3841 [Erratum-ibid.\ A {\bf
9} (1994) 4353] [arXiv:hep-th/9306002].
%
%
\bibitem{Sasaki:1993xr}
R.~Sasaki, {\em `Reflection bootstrap equations for Toda field
  theory',}
in the proceedings of the conference, Interface between physics and
mathematics, eds.\ W.~Nahm and J.-M.\ Shen
[arXiv:hep-th/9311027].
%
\bibitem{doreyup}
P.~Dorey, unpublished.
%
\bibitem{Corrigan:1994ft}
E.~Corrigan, P.~Dorey, R.H.~Rietdijk and R.~Sasaki, {\em `Affine
Toda field theory on a half line',} Phys.\ Lett.\ B {\bf 333} (1994)
83 [arXiv:hep-th/9404108].
%
\bibitem{Corrigan:1994np}
E.~Corrigan, P.~Dorey and R.H.~Rietdijk, {\em `Aspects of affine
Toda field theory on a half line',} Prog.\ Theor.\ Phys.\ Suppl.\
{\bf 118} (1995) 143 [arXiv:hep-th/9407148].
%
%
\bibitem{Fateev:2001mj}
V.A.~Fateev, {\em `Normalization factors, reflection amplitudes and
integrable systems',} [arXiv:hep-th/0103014].
%
\bibitem{cz}
C.~Zambon, York PhD Thesis 2004, unpublished.
%
\bibitem{Dorey:1992bq}
P.~Dorey and F.~Ravanini, {\em `Staircase models from affine Toda
field theory',} Int.\ J.\ Mod.\ Phys.\ A {\bf 8} (1993) 873
[arXiv:hep-th/9206052].
%
%
\bibitem{Ghoshal:1993iq}
S.~Ghoshal, {\em `Bound state boundary S matrix of the sine-Gordon
model',} Int.\ J.\ Mod.\ Phys.\ A {\bf 9} (1994) 4801
[arXiv:hep-th/9310188].
%
%
\bibitem{deVega:1992zd}
H.J.~de Vega and A.~Gonzalez Ruiz, {\em `Boundary K matrices for the
six vertex and the n(2n-1) A(n-1) vertex models',} J.\ Phys.\ A {\bf
26} (1993) L519 [arXiv:hep-th/9211114].
%
%
\bibitem{Cardy:1984bb}
J.L.~Cardy, {\em `Conformal invariance and surface critical
behavior',} Nucl.\ Phys.\ B {\bf 240} (1984) 514.
%
\bibitem{Zamolodchikov:1991et}
Al.B.~Zamolodchikov, {\em `On the thermodynamic Bethe ansatz
equations for reflectionless ADE scattering theories',} Phys.\
Lett.\ B {\bf 253} (1991) 391.
%
\bibitem{Fendley:1994rh}
P.~Fendley, H.~Saleur and N.P.~Warner, {\em `Exact solution of a
massless scalar field with a relevant boundary interaction'}, Nucl.\
Phys.\ B {\bf 430}, (1994) 577 [arXiv:hep-th/9406125].
%
\bibitem{Bazhanov:1994ft}
V.V.~Bazhanov, S.L.~Lukyanov and A.B.~Zamolodchikov, {\em
`Integrable structure of conformal field theory, quantum KdV theory
and thermodynamic Bethe ansatz'}, Commun.\ Math.\ Phys.\  {\bf 177},
(1996) 381  [arXiv:hep-th/9412229].
%
\bibitem{Bazhanov:2001xm}
V.V.~Bazhanov, A.N.~Hibberd and S.M.~Khoroshkin, {\em `Integrable
structure of W(3) conformal field theory, quantum Boussinesq theory
and boundary affine Toda theory'}, Nucl.\ Phys.\ B {\bf 622},
(2002)475 [arXiv:hep-th/0105177].
%
\bibitem{Smirnov:1990vm}
F.A.~Smirnov, {\em `Reductions of the sine-Gordon model as a
perturbation of minimal models of conformal field theory',} Nucl.\
Phys.\ B {\bf 337} (1990) 156.
%
%
\bibitem{Chatterjee:1995be}
R.~Chatterjee, {\em `Exact partition function and boundary state of
2-D massive Ising field theory with boundary magnetic field',}
Nucl.\ Phys.\ B {\bf 468} (1996) 439 [arXiv:hep-th/9509071].
%
\bibitem{Fateev:1997yg}
V.~Fateev, S.L.~Lukyanov, A.B.~Zamolodchikov and
Al.B.~Zamolodchikov,
{\em `Expectation values of local fields in Bullough-Dodd model and
integrable perturbed conformal field theories',}
Nucl.\ Phys.\ B {\bf 516} (1998) 652
[arXiv:hep-th/9709034].
%
\bibitem{Guida:1997fs}
R.~Guida and N.~Magnoli,
{\em `Vacuum expectation values from a variational approach',}
Phys.\ Lett.\ B {\bf 411} (1997) 127
[arXiv:hep-th/9706017].
%
\bibitem{DiFrancesco:1997nk}
P.~Di Francesco, P.~Mathieu and D.~Senechal, {\em `Conformal field
theory',} Springer, 1997.
%
\bibitem{Goddard:1984vk}
P.~Goddard, A.~Kent and D.I.~Olive, {\em `Virasoro algebras and
coset space models',} Phys.\ Lett.\ B {\bf 152} (1985) 88.
%
\bibitem{Gannon:2001py}
T.~Gannon, {\em`Algorithms for affine Kac-Moody algebras',}
[arXiv:hep-th/0106123].
%
\bibitem{Schell}
B.~Schellekens, {\em
http://www.nikhef.nl/{\textasciitilde}t58/kac.html}.
%
\bibitem{Affleck:1998nq}
I.~Affleck, M.~Oshikawa and H.~Saleur, {\em `Boundary critical
phenomena in the three-state Potts model',}
[arXiv:cond-mat/9804117].
%
\bibitem{Chim:1995kf}
L.~Chim, {\em `Boundary S-matrix for the tricritical Ising model',}
Int.\ J.\ Mod.\ Phys.\ A {\bf 11} (1996) 4491
[arXiv:hep-th/9510008].
%
\bibitem{Fateev:1993av}
V.A.~Fateev, {\em `The exact relations between the coupling
constants and the masses of particles for the integrable perturbed
conformal field theories',} Phys.\ Lett.\ B {\bf 324} (1994) 45.
%
\bibitem{Fredenhagen:2003xf}
S.~Fredenhagen, {\em `Organizing boundary RG flows',} Nucl.\ Phys.\
B {\bf 660} (2003) 436 [arXiv:hep-th/0301229].
%
\bibitem{Affleck:2000jv}
I.~Affleck, {\em `Edge critical behaviour of the 2-dimensional
tri-critical Ising model',} J.\ Phys.\ A {\bf 33} (2000) 6473
[arXiv:cond-mat/0005286].
%
\bibitem{Graham:2001pp}
  K.~Graham,
  {\em `On perturbations of unitary minimal models by boundary condition
 changing operators'},
  JHEP {\bf 0203}, 028 (2002)
  [arXiv:hep-th/0111205].
  %
\bibitem{Cardy:1989ir}
J.L.~Cardy, {\em  `Boundary conditions, fusion rules and the
Verlinde formula',}
  Nucl.\ Phys.\ B {\bf 324} (1989) 581.
%
\bibitem{Saleur:1988zx}
H.~Saleur and M.~Bauer, {\em `On some relations between local height
probabilities and conformal invariance',}
  Nucl.\ Phys.\ B {\bf 320} (1989) 591.
%
\bibitem{Gepner:1989jq}
D.~Gepner, {\em `Field identification in coset conformal field
theories',} Phys.\ Lett.\ B {\bf 222} (1989) 207.
%
\bibitem{Schellekens:1989uf}
A.N.~Schellekens and S.~Yankielowicz, {\em `Field identification
fixed points in the coset construction',} Nucl.\ Phys.\ B {\bf 334}
(1990) 67.
%
%
\bibitem{Ahn:1989re}
C.R.~Ahn and M.A.~Walton, {\em `Field identifications in coset
conformal theories from projection matrices',} Phys.\ Rev.\ D {\bf
41} (1990) 2558.
%
%
\bibitem{Frenkel:1980rn}
I.B.~Frenkel and V.G.~Kac, {\em `Basic representations of affine Lie
algebras and dual resonance models',} Invent.\ Math.\  {\bf 62}
(1980) 23.
%
%
\bibitem{Kac:1984mq}
V.G.~Kac and D.H.~Peterson, {\em `Infinite dimensional Lie algebras,
theta functions and modular forms',} Adv.\ Math.\  {\bf 53} (1984)
125.
%
\bibitem{Kac:1990gs}
V.G.~Kac, {\em `Infinite dimensional Lie algebras',} Cambridge
University Press, 1990.
%
\bibitem{Bernard:1986xy}
D.~Bernard, {\em `String characters from Kac-Moody automorphisms',}
Nucl.\ Phys.\ B {\bf 288} (1987) 628.
%
\bibitem{Fuchs:1990wb}
J.~Fuchs, {\em `Simple WZW currents',} Commun.\ Math.\ Phys.\  {\bf
136} (1991) 345.
%
\end{thebibliography}
\end{document}